\documentstyle[12pt]{article}
\def\three_j(#1,#2,#3,#4,#5,#6){\pmatrix{#1 & #2 & #3\cr
                                         #4 & #5 & #6\cr}}

\def\qqq{\end{document}}
\def\pmb#1{\setbox0=\hbox{$#1$}%
\kern-.025em\copy0\kern-\wd0
\kern.05em\copy0\kern-\wd0
\kern-.025em\raise.0433em\box0 }

\def\xara(#1,#2,#3,#4){\left(\matrix{#1 & #2\cr #3 & #4\cr}\right)}

\def\w{\omega}

\def\six_j(#1,#2,#3,#4,#5,#6){\left\{\matrix{#1 & #2 & #3\cr
                                         #4 & #5 & #6\cr}\right\}}
\def\nine_j(#1,#2,#3,#4,#5,#6,#7,#8,#9){\left\{\matrix{#1 & #2 & #3\cr
                                        #4 & #5 & #6\cr
                                         #7 & #8 & #9\cr}\right\}}

\def\Ener(#1,#2){ \sqrt{{#1}^2+{#2}^2} }

\def\overlay#1#2{\setbox0=\hbox{$#1$}\setbox1=\hbox to \wd0{\hss$#2$\hss}#1%
\hskip -1\wd0\copy1}
\newcommand{\xslash}[1]{\overlay{#1}{/}}
\newcommand{\sla}[1]{\xslash{#1}}

\def\bold#1{\setbox0=\hbox{$#1$}%
      \kern-.025em\copy0\kern-\wd0
      \kern.05em\copy0\kern-\wd0
      \kern-.025em\raise.0433em\box0 }

\def\Tr{\, \hbox{Tr} \, }

\def\gsim{\displaystyle\mathop{>}_{\sim}}
\def\lsim{\displaystyle\mathop{<}_{\sim}}

\def\S11{S_{11}(1535)}
\def\E0+{E_{0^+}}

\def\footnoterule{\kern-3pt \hrule width \hsize \kern2.6pt}

\newcommand{\beq}{\begin{equation}}
\newcommand{\eeq}{\end{equation}}
\newcommand{\ba}{\begin{array}}
\newcommand{\ea}{\end{array}}
\newcommand{\beqa}{\begin{eqnarray}}
\newcommand{\eeqa}{\end{eqnarray}}
\newcommand{\bd}[1]{ \mbox{\boldmath $#1$}  }

\begin{document}
\setcounter{footnote}{0}
\begin{center}
{\Large \bf Analysis of factorization in (e,e$'$p) reactions:
A survey of the relativistic plane wave impulse approximation}\\
\vspace*{1cm}
J.A. Caballero$^{1,2}$, T.W. Donnelly$^{3}$, E. Moya de Guerra$^{2}$ and 
J.M. Ud\'{\i}as$^{4}$\\
\vspace*{0.7cm}
$^{1}${\sl Departamento de F\'{\i}sica At\'omica,
Molecular y Nuclear \\ 
Universidad de Sevilla, Apdo: 1065, Sevilla 41080, SPAIN }\\
\vspace{0.35cm}
$^{2}${\sl Instituto de Estructura de la Materia, 
	Consejo Superior de Investigaciones Cient\'{\i}ficas  \\ 
	 Serrano 123, Madrid 28006, SPAIN }\\
\vspace*{0.35cm}
$^{3}${\sl Center for Theoretical Physics, Laboratory for Nuclear Science and
	Dept. of Physics\\
	Massachusetts Institute of Technology, 
	Cambridge, MA 02139--4307, USA }\\
\vspace{0.35cm}
$^{4}${\sl Departamento de F\'{\i}sica At\'omica, Molecular y Nuclear \\
	Universidad Complutense de Madrid, Avda. Complutense s/n, 
	Madrid 28040, SPAIN} 
\end{center}
\vspace{0.75cm}
\begin{abstract}
The issue of factorization within the context of coincidence
quasi--elastic electron scattering is reviewed.
Using a relativistic formalism for the entire reaction mechanism 
and restricting ourselves to the case of plane waves for the
outgoing proton, we
discuss the meaning of factorization in the cross section and the
role of the small components of the bound nucleon
wave function.
\end{abstract}
\vspace{0.8cm}
PACS number(s): 25.30.Fj, 24.10.Jv, 21.10.Jx 
\vfill
\eject
\section*{1. Introduction}

As is well known~\cite{fm85} in standard plane--wave impulse approximation (PWIA)
the differential cross section for A(e, e$'$p)B reactions factorizes into an
elementary cross section $\sigma _{ep}$, describing electron proton
scattering, and a spectral function $S(E_m,p_m)$, describing the probability
for finding a proton in the target nucleus with energy $(E_m)$ and
momentum $(p_m)$ compatible with the kinematics of the process. The
differential cross section in PWIA is thus written as 
\begin{equation}
\label{(1)}\frac{d^6\sigma }{d\varepsilon' d\Omega _{e}dE_p d\Omega _p}=\chi
\sigma _{ep}\ S(E_m,p_m)
\label{int1}
\end{equation}
with $\chi $ a kinematical factor.
It is this factorization property that makes A(e, e$'$p)B reactions so
appealing for investigations of nuclear structure.

According to eq.~(\ref{int1})
single--particle distributions can in principle be probed in great detail. In
fact several orbitals have been mapped out in many nuclei over rather
extended momentum ranges~\cite{hu90}--\cite{bo94}, in spite of the limitations of the PWIA
approximation. These limitations are also well known. Final--state
interactions and, more generally, distortion of both electron and nucleon
wave functions due to electromagnetic and strong interactions with target
and residual nuclei destroy the elegant simplicity of eq.~(\ref{int1}).

Nevertheless this equation is still very useful and is the basis for
interpretation of experimental data. The latter are usually compared to
theory defining an effective spectral function 
\begin{equation}
\label{(2)}\stackrel{\sim }{S}\ (E_m,p_m)=\left( \chi \sigma _{ep}\right)
^{-1}\frac{d^6\sigma }{d\varepsilon' d\Omega _{e\ }dE_p\
d\Omega _p}
\label{int2}
\end{equation}
and an effective proton momentum distribution (or reduced cross 
section~\cite{hu90,qu88}) 
\begin{equation}
\label{(3)}\stackrel{\sim }{\rho }\ (p_m)=\int_{\Delta E_m}\stackrel{\sim }{S%
}\ (E_m,p_m)\ dE_m,
\label{int3}
\end{equation}
that in PWIA corresponds
exactly to the density in momentum space of the single--particle orbital
associated with a selected value of the missing energy
$E_m$. In general $\stackrel{\sim }{S}$ and
$\stackrel{\sim }{\rho}$ depend not only on $p_m$ and $E_m$ but also on the
remaining kinematical variables.

Comparisons of theories taking into account in various ways final--state
interactions and/or Coulomb distortions are also made for this effective
momentum distribution. When final--state interactions and Coulomb distortions
are fully taken into account, departures between experiment and theory can
solely be attributed to limitations of the single--particle nuclear model
and impulse approximation. Thus eqs.~(\ref{int1}) 
to~(\ref{int3}) form also a basis for the empirical study of
nucleon--nucleon correlations in the target. 

In recent years much work has
been done along these lines on both experimental and theoretical fronts.
Experimental exclusive (e, e$'$p) measurements have been made with high
precision on quite extensive missing momentum 
regions~\cite{hu90}--\cite{bo94}. The analyses
of data based on standard distorted--wave impulse approximation (DWIA) have
met two major difficulties. On the one hand the spectroscopic factors
extracted from DWIA analyses of low p$_m$ $(p_m\leq 300\ $%
MeV$)$ data are too small compared with theoretical predictions. As an example
the extracted occupations of $3s_{1/2}$ and $2d_{5/2}$ orbits in $^{208}$%
Pb are $S_\alpha \simeq 0.5$, while theories on short--range 
correlations~\cite{pa84}
predict at most a 30 \% reduction of mean field occupations for levels just
below the Fermi level. On the other hand the high-$p_m$ data $(300\ $MeV$\
\leq p_m<600\ $MeV$)$ on $\stackrel{\sim}{\rho}(p_m)$ for the same
levels are much larger than the results of DWIA calculations 
that are compatible with
those that fit the low-$p_m$ data~\cite{bo94}. Although short--range correlations
are expected to increase the high momentum components, their effect is
negligible~\cite{mu94} at the small missing energies of the existing high-$p_m$
data, and effects of long--range correlations have been invoked~\cite{bo94}. The
above mentioned DWIA\ analyses are based on non--relativistic independent
particle models and use phenomenological (real and imaginary) Woods-Saxon
potentials with parameters fitted to individual nuclei and shells.

Alternatively, in recent years, the relativistic mean field approximation
has been successfully used for the analyses of both low- and high-$p_m$ 
data~\cite{mc90}--\cite{ud96}. In the relativistic distorted--wave 
impulse approximation (RDWIA)
bound and scattered nucleons are described by solutions of the Dirac
equation with scalar and vector (S-V) potentials. In the work 
of refs.~\cite{ud93}--\cite{ud96}
the scattering nucleon wave functions were obtained solving the Dirac
equation with the relativistic optical potentials of ref.~\cite{ha90} 
and the bound
nucleon wave functions were obtained with the TIMORA code~\cite{ho91} based on
Walecka's mean field approximation~\cite{se86}. The parameters of the S-V
potentials are obtained from global fits and, at variance with DWIA, 
in RDWIA there
were no parameters fitted to the particular levels other than the
spectroscopic factors. In all the cases studied the extracted spectroscopic
factors are larger than in DWIA and are valid for low- and high-$p_m$ 
data~\cite{mc90}--\cite{ud96}. 
For instance, for the above mentioned $3s_{1/2}$ and $2d_{3/2}$
shells in $^{208}$Pb values of $S_\alpha \simeq 0.7$ have been 
obtained~\cite{mc90,ud93},
consistent with theoretical predictions, and a reasonable agreement
between RDWIA calculations and experiment has been found in the high-$p_m$
region~\cite{ud96}.

A global comparison of the unscaled results shows that, compared to DWIA,
the reduced cross sections obtained in RDWIA are smaller at low $p_m$ and
are larger in the high-$p_m$ region, thus providing a consistent description
of all available data with moderate values of the spectroscopic factors.

The reasons why RDWIA produce smaller cross sections in the low-$p_m$ region
were investigated in refs.~\cite{ud95,ji94}. In ref.~\cite{ud95} 
the success of the
relativistic analyses was traced back to the improved treatment of
distortion effects in the electron and outgoing proton waves. In the high $%
p_m$ region other factors can be important. In particular, the presence of
higher momentum components in the relativistic bound nucleon wave function,
and in the relativistic nucleon current operator can be expected to play a
role in this region. In view of the success of RDWIA calculations it is
worthwhile investigating more deeply the dependence of the results on the lower
components of the bound nucleon wave function and on the choice of the
current operator. This is the purpose of this paper where we restrict
ourselves to the relativistic plane--wave impulse approximation (RPWIA).

The reasons why we consider here only the plane--wave limit are twofold. On
one hand it is convenient to restrict the study to RPWIA to be able to
disentangle effects of distortion from effects of high $p_m$ components in
the bound nucleon wave function and current operator. On the other hand it
is interesting to study this approximation because an important difference
between relativistic and non-relativistic approaches already appears in the
plane--wave limit. Namely, the factorized expression in eq.~(\ref{int1}) does not
necessarily hold in RPWIA. As we shall see in detail, this difference comes
from the fact that in the non-relativistic approximation the nuclear current
is expanded on a basis of free (positive energy) Dirac spinors, while in the
relativistic approximation the nuclear current is directly written in terms
of the relativistic bound nucleon wave function that contains both positive
and negative energy projections in the complete Dirac basis. Thus the
question of whether factorization (in the sense of eq.~(\ref{int1})) does or does not
hold in RPWIA is intimately connected with the role of the negative energy
projection of the bound nucleon wave function.

Although the physical meaning of contributions from negative energy
projections can be (and indeed is) a matter of debate~\cite{br84}, and their
presence may be questioned, we would like to stress that in this respect our
aim here is to analyze the quantitative importance of those contributions
rather than to take sides on a particular conceptual interpretation or
criterion.

As we shall see the extent to which negative energy components are important
depends on the choice of the current operator. Thus the present analyses is
also useful in choosing a particular form of the current 
operator once one adhers to a
given criterion. For instance if one would like to follow the point of view
that negative energy projections should not contribute to physical
observables one would choose a current operator that minimizes the role of
such components.

The organization of the paper is as follows.
In section 2 we discuss in detail the relationship between RPWIA and PWIA 
differential cross sections. A separation of the RPWIA cross section and response
functions is made into positive and negative energy projections. In section 3
we discuss various choices of the nuclear current operator and their influence
on the positive and negative energy projections of the single--nucleon responses
and cross sections. In section 4 we analyze the role of the negative energy
projections and the lack of factorization in the total nuclear responses and
differential cross sections. The main conclusions are summarized in section 5.

\section*{2. Formal comparison of RPWIA to PWIA}

\subsection*{2.1 Differential cross section in RPWIA}

The general formalism for exclusive electron scattering reactions has been
presented in detail in several previous papers.
We refer in particular to refs.~\cite{fm85} and~\cite{ud93} for the
nonrelativistic and relativistic treatments, respectively.
Here we just summarize the kinematics and focus on those aspects 
of the cross section that are of
relevance to the points of discussion in this paper.

In the (e,e$'$p) process an electron with four
momentum $K^\mu=(\varepsilon,{\bd k})$ is scattered through an angle
$\theta_e$ to four momentum $K'^\mu=(\varepsilon',{\bd k'})$. 
We denote the hadronic variables by $P_A^\mu=(M_A,{\bd O})$, 
$P_B^\mu=(E_B,{\bd p}_B)$, $P_N^\mu=(E_N,{\bd p}_N)$ the four momenta of
the target, residual nucleus and outgoing proton, respectively.
The target rest mass is $M_A$, and $M=\sqrt{E_N^2-{\bd p}_N^2}$,
$M_B=\sqrt{E_B^2-{\bd p}_B^2}$ are 
the outgoing proton and residual nucleus masses 
(the latter may include possible internal excitation).
As usual in these processes electrons are treated in the extreme
relativistic limit (ERL): $\varepsilon =k$, $\varepsilon' =k'$.
The four momentum transfer is given by
$Q^\mu=K^\mu-K'^\mu =(\omega,{\bd q})$.
For the applications discussed in the next sections we consider $^{16}$O and
have selected three kinematic situations:
\begin{itemize}
\item I) $q=500$ MeV/c, $\w=131.56$ MeV
\item II) $q=1$ GeV/c, $\w=432.8$ MeV 
\item III) $q=1$ GeV/c, $\w=300$ MeV. 
\end{itemize}
The value of $\omega$ in kinematics I and
II corresponds to the quasielastic peak value ($\omega_{QE}$) while in
kinematics III, $\omega < \omega_{QE}$, the so--called y--scaling region.
Denoting by ${\bd p}$ the three momentum of the struck nucleon 
(${\bd p}={\bd p}_m$, the missing momentum, in the RPWIA, as discussed below),
the range of variation of $p$--values has been chosen $0\leq p\leq 500$
MeV/c. From similar arguments to those presented in ref.~\cite{edu} 
it can be shown that the
variation of the momentum $p_N$ is negligible, since one has that
$M_B^2\gg p^2$.
In fact, for $A=16$ this variation is of the order of $\sim 4\%$ 
(kinematics I) and $\sim 1.5\%$ (kinematics II, III) in the
whole range of $p$--values considered. 

In order to compare the content of the relativistic and non--relativistic
treatments in plane--wave impulse approximation we focus, without lack of
generality, on the differential cross section leading to a specific final
state of the residual nucleus, corresponding to proton knock--out from a
specific bound orbital $b$ in the target nucleus. Following ref.~[8], and
taking unity spectroscopic factors and plane waves for incoming electron
and for outgoing electron and proton, we write the differential cross
section in RPWIA as

\beq
\frac{d^5\sigma}{d\Omega_ed\varepsilon' d\Omega_N}=
\frac{2\alpha^2}{Q^4}\left(\frac{\varepsilon'}{\varepsilon}\right)
\frac{p_N M M_B}{M_A f_{rec}} 2
 	\overline{\sum}|{\cal M}|^2 \,\,\, ,
\label{1}
\eeq
where $\overline{\sum}$ denotes average over initial and sum over final
polarizations. The transition amplitude is the contraction of the electron
($j^\mu _e$) and nucleon ($J^\mu _N$) currents
\beq
{\cal M}=j^e_\mu J^\mu _N
\label{e1}
\eeq
with
\beqa
j^\mu_e &=& \overline{u}^e_{\sigma _f}({\bd k}')\gamma^\mu 
u^e_{\sigma_i}({\bd k}) \label{e3a} \\
J^\mu_N &=& \overline{u}_{\sigma _N}({\bd p}_N)\hat{J}^\mu _N 
\Psi_b^{m_b}({\bd p}) \label{e3b} \,\,\, ,
\eeqa
where $\hat{J}^\mu _N$ is the nucleon current operator to be discussed in
section 3 and the momentum and energy of the bound nucleon satisfy,
${\bd p}={\bd p}_N -{\bd q} =-{\bd p}_B$; 
$\epsilon_b=-M_A+M_B+M=\omega-T_N-T_B$. The term 
$\Psi_b^{m_b}({\bd p})$ denotes the Fourier transform of the relativistic
bound nucleon wave function
\beq
\Psi_b^{m_b}({\bd p})=\frac{1}{(2\pi)^{3/2}}\int d{\bd p} 
e^{-i{\bd p}\cdot {\bd r}}\Psi_b^{m_b}({\bd r})
\label{e4}
\eeq
with $\Psi_b^{m_b}({\bd r})$ a solution of the Dirac equation with S--V
potentials. For details see appendix A and refs.~\cite{ho91,se86}.

\subsection*{2.2 Positive and negative energy projections of the bound
nucleon wave function and nucleon current}

It is known that only free solutions can be expanded in terms of free
positive--energy Dirac spinors $u$ alone. For an interacting
(i.e., not free) relativistic wave
function there is always a coupling to the free negative--energy Dirac
spinors $v$. This implies that the free relation between upper ($u$) and lower
($d$) components
\beq
\frac{{\bd \sigma} \cdot {\bd p}}{\overline{E}+M}\phi^{u} = \phi^{d}
\label{11}
\eeq
does not hold in general for a bound relativistic wave function. As we
shall see in detail, this is what differentiates RPWIA from PWIA. To
analyze this point we insert in eq.~(\ref{e3b})
the completeness relation~\cite{bd}
\beq
\sum_s\left[ u_\alpha ({\bd p},s)\overline{u}_\beta({\bd p},s)-
		v_\alpha ({\bd p},s)\overline{v}_\beta({\bd p},s)\right]
	=\delta_{\alpha \beta}
\label{e10}
\eeq
and write
\beq
J^\mu_{N}= \langle J^\mu _N \rangle_u - \langle J^\mu _N \rangle_v
\,\,\, ,
\label{e11}
\eeq
where the first term, indicated by the index $u$, comes from the 
positive--energy projector involving the Dirac
spinors $u({\bd p},s)$ 
and the second term, indicated by the index
$v$ comes from the negative--energy projector involving the Dirac
spinors $v({\bd p},s)$.
These contributions are given by
\beq
\langle J^\mu _N \rangle_u \equiv
\langle J^\mu \rangle_u=\sum_s\overline{u}({\bd p}_N,s_N)
\hat{J}^\mu u({\bd p},s)\left[\overline{u}({\bd p},s)
		\Psi_b^{m_b}({\bd p})\right]
\label{15}
\eeq

\beq
\langle J^\mu _N \rangle_v \equiv
\langle J^\mu \rangle_v=\sum_s\overline{u}({\bd p}_N,s_N)
\hat{J}^\mu v({\bd p},s)\left[\overline{v}({\bd p},s)
		\Psi_b^{m_b}({\bd p})\right]
\label{16} \,\,\, .
\eeq
To simplify the notation, here and in what follows we suppress the index
$N$ on nucleon 
currents and current operators.

Using the explicit expression of the relativistic
bound nucleon wave function in momentum space
given in appendix A, the $u$ and $v$ contractions in equations~(\ref{15}) 
and~(\ref{16}) are found to be
\beq
\left[\overline{u}({\bd p},s)\Psi_b^{m_b}({\bd p})\right]=
(-i)^\ell\sqrt{\frac{
\overline{E}+M}{2M}}\alpha_{\kappa_b}(p)
\langle s |\Phi_{\kappa_b}^{m_b} (\hat{{\bd p}}) \rangle
\label{18}
\eeq

\beq
\left[\overline{v}({\bd p},s)\Psi_b^{m_b}({\bd p})\right]= 
	(-i)^\ell\sqrt{\frac{\overline{E}+M}{2M}}\beta_{\kappa_b}(p)
	\langle s |\Phi_{-\kappa_b}^{m_b} (\hat{{\bd p}}) \rangle
\,\,\, , \label{21}
\eeq
where $\langle s |\Phi_{\pm \kappa_b}^{m_b} (\hat{{\bd p}}) \rangle$
indicates spin projections of the bispinors 
$\Phi_{\pm \kappa_b}^{m_b}$ on a spin state $|\frac{1}{2} s\rangle$, and the
radial functions in momentum space $\alpha_{\kappa_b}$ and $\beta_{\kappa_b}$ 
are given 
by
\beqa
\alpha_{\kappa_b}(p)&=&g_{\kappa_b}(p)-
\frac{p}{\overline{E}+M}S_{\kappa_b} f_{\kappa_b}(p)
\label{19} \\
\beta_{\kappa_b}(p)&=&\frac{p}{\overline{E}+M}g_{\kappa_b}(p)-
S_{\kappa_b} f_{\kappa_b}(p)
\label{22}
\eeqa
with $g_{\kappa_b}$ and $f_{\kappa_b}$ the Bessel transforms 
of the standard upper
and lower radial functions of the 
bound nucleon wave function in coordinate space
(see appendix A for details).

Obviously for a free nucleon
$\beta_{\kappa}(p)=0$. For a bound nucleon $\beta_{\kappa_b} (p) \neq 0$
in general,
unless we impose the relation in eq.~(\ref{11}) 
which amounts to imposing $\beta_{\kappa_b}(p)=0$.

\subsection*{2.3 Tensorial decomposition}
 
In order to connect with the standard PWIA we perform the usual
decomposition of the transition probability in eq.~(\ref{1}) 
into a leptonic and a hadronic tensor
\beq
2\overline{\sum}|{\cal M}|^2 = \eta_{\mu \nu}W^{\mu\nu} \,\,\, ,
\label{e5}
\eeq
where $\eta_{\mu\nu}$ is the leptonic tensor
\beq
\eta_{\mu\nu}=\sum_{\sigma_i \sigma_f} j^e_\mu j_\nu^{e\ast}
\label{e6}
\eeq
whose explicit expression can be found for instance in 
refs.~\cite{fm85,don}, and $W^{\mu\nu}$ is the hadronic tensor 
\beq
W^{\mu\nu}=\frac{2}{2j_b+1}\sum_{s_N m_b}J^\mu J^{\nu\ast}\,\,\, .
\label{e7}
\eeq
We recall that in standard PWIA the hadronic tensor factorizes into a
tensor for elastic scattering on a free proton ${\cal W}^{\mu\nu}$ and the
momentum distribution of the (non--relativistic) bound orbital $N_b(p)$
\beq
\left(W^{\mu\nu}\right)_{PWIA} = {\cal W}^{\mu\nu}({\bd p},{\bd q})
N_b(p)
\label{e8}
\eeq
with
\beq
{\cal W}^{\mu\nu}({\bd p};{\bd q})
	=\sum_{s,s_N}\left[\overline{u}({\bd p}_N,s_N)
	\hat{J}^\mu u({\bd p},s)\right]^*
	\left[\overline{u}({\bd p}_N,s_N)
	\hat{J}^\nu u({\bd p},s)\right]
\label{9}
\eeq
and
\beq
N_b(p)=\frac{1}{2j_b+1}\sum_{m_b}\left|\phi_b^{m_b}({\bd p}) \right|^2 
\,\,\, ,
\label{10}
\eeq
where $\phi_b^{m_b}({\bd p})$ is the (non--relativistic) wave function of
the bound nucleon in momentum space (see for instance 
refs.~\cite{fm85,edu,ann}).

The connection between RPWIA and PWIA can be best seen by substitution of
eqs.~(\ref{e11}-\ref{21}) into eq.~(\ref{e7}). This leads to the following expression for
the hadronic tensor
\beq
W^{\mu\nu}=W^{\mu\nu}_P+W^{\mu\nu}_N+W^{\mu\nu}_C \,\,\, ,
\label{24}
\eeq
where $W^{\mu\nu}_P$ ($W^{\mu\nu}_N$) is the contribution from positive
(negative) energy projections only, while $W^{\mu\nu}_C$ is a crossed term
containing products of both positive and negative energy projections.
These components of the hadronic tensor
are given explicitly as follows
\beq
W^{\mu\nu}_P=
\frac{2}{2j_b+1}\sum_{m_b}\sum_{s_N}\langle\hat{J}^\mu \rangle^*_u
	\langle\hat{J}^\nu \rangle_u
\label{29p}
\eeq
\beq
W^{\mu\nu}_N=
\frac{2}{2j_b+1}\sum_{m_b}\sum_{s_N}\langle\hat{J}^\mu \rangle^*_v
	\langle\hat{J}^\nu \rangle_v
\label{36p}
\eeq
\beq
W^{\mu\nu}_C=
\frac{-2}{2j_b+1}\sum_{m_b}\sum_{s_N}\left(
	\langle \hat{J}^\mu \rangle_u^\ast\langle \hat{J}^\nu \rangle _v+
	\langle \hat{J}^\mu \rangle_v^\ast\langle \hat{J}^\nu \rangle _u
	\right) \,\,\, .
\label{63p}
\eeq

Using the relations
\beqa
\sum_{m_b} \langle s |\Phi_{\kappa_b}^{m_b} \rangle^\ast
 	\langle s'|\Phi_{\kappa_b}^{m_b} \rangle &=&
	\delta_{ss'}\frac{2j_b+1}{8\pi}
\label{62pp} \\
\sum_{m_b} \langle s |\Phi_{\kappa_b}^{m_b} \rangle^\ast
 	\langle s'|\Phi_{-\kappa_b}^{m_b} \rangle &=&
	-\langle s'|{\bd \sigma}\cdot {\bd p}|s\rangle\frac{2j_b+1}{8p\pi} 
\label{63pp}
\eeqa
it is easy to carry out the sum over $m_b$ in 
eqs.~(\ref{29p}-\ref{63p}) to get (see appendix B)
\beq
W^{\mu\nu}_P=
	N_{uu}(p){\cal W}^{\mu\nu}
\label{29pc}
\eeq
\beq
W^{\mu\nu}_N=
	N_{vv}(p){\cal Z}^{\mu\nu}
\label{36pc}
\eeq
\beq
W^{\mu\nu}_C=
	N_{uv}(p){\cal N}^{\mu\nu} \,\,\, ,
\label{63pc}
\eeq
where we use the fact that in our case $\alpha_{\kappa_b}(p)$ and
$\beta_{\kappa_b}(p)$ are real to write
\beqa
N_{uu}(p) &=& \left(\tilde{\alpha}_{\kappa_b} (p)\right)^2
	\label{77p1} \\
N_{vv}(p) &=& \left(\tilde{\beta}_{\kappa_b}(p)\right)^2
	\label{77p2} \\
N_{uv}(p) &=& -2\tilde{\alpha}_{\kappa_b}(p)\tilde{\beta}_{\kappa_b}(p)
	\label{77p3}
\eeqa
with
\beqa
\tilde{\alpha}_{\kappa_b} (p) &=&
	\sqrt{\frac{\overline{E}+M}{8\pi M}}\alpha_{\kappa_b} (p)
		\label{77pc} \\
\tilde{\beta}_{\kappa_b} (p) &=&
	\sqrt{\frac{\overline{E}+M}{8\pi M}}\beta_{\kappa_b} (p)
		\label{78pc}
\eeqa
and the single--nucleon tensors ${\cal W}^{\mu\nu}$, ${\cal Z}^{\mu\nu}$,
${\cal N}^{\mu\nu}$ are given by
\beq
{\cal W}^{\mu\nu}=
\Tr \left[\left(\frac{\sla{P}+M}{2M}\right)\overline{J}^\mu
	\left(\frac{\sla{P}_N+M}{2M}\right)J^\nu
	\right]
\label{30}
\eeq

\beq
{\cal Z}^{\mu\nu}=
\Tr \left[\left(\frac{\sla{P}-M}{2M}\right)\overline{J}^\mu
	\left(\frac{\sla{P}_N+M}{2M}\right)J^\nu
	\right]
\label{37p}
\eeq

\beq
{\cal N}^{\mu\nu}=
\Tr \left[\overline{J}^\mu\left(\frac{\sla{P}_N+M}{2M}\right)
	J^\nu\gamma^0\frac{{\bd \gamma}\cdot{\bd p}}{p}
	\frac{\sla{P}}{2M}
	\right] \,\,\, ,
\label{37pp}
\eeq
where we use the notation $\overline{J}^\mu\equiv \gamma_0J^{\mu+}\gamma_0$. 
Explicit expressions for these tensors for different current operators are
given in appendix C.

As already mentioned ${\cal W}^{\mu\nu}$ is the usual single--nucleon tensor
appearing in the standard PWIA for unpolarized scattering, while
${\cal Z}^{\mu\nu}$ and ${\cal N}^{\mu\nu}$ are new single--nucleon
tensors that appear only when the bound nucleon wave functions contain a
non--zero negative--energy projection. Using the identity
\beq
{\bd \gamma}\cdot {\bd p}= {\bd \sigma}\cdot {\bd p} \gamma_0 \gamma_5
\label{n197}
\eeq
and following the arguments given in ref.~\cite{dp}
in the context of PWIA with polarized beam and target, it can be also shown
that the single--nucleon tensor ${\cal N}^{\mu\nu}$ can be related to
a diagonal tensor constructed from spinors quantized with respect to
a spin axis pointing along a generic direction,
${\cal R}^{\mu\nu}(\theta_R,\phi_R)$, as
\beq
{\cal N}^{\mu\nu}\equiv
	\cos\theta{\cal R}^{\mu\nu}(0,0)+
	\sin\theta\left(\cos\phi{\cal R}^{\mu\nu}(\frac{\pi}{2},0)
	+\sin\phi{\cal R}^{\mu\nu}(\frac{\pi}{2},\frac{\pi}{2})
	\right)
\label{75a}
\eeq
with $\theta$, $\phi$ defining the direction of the bound nucleon momentum
${\bd p}$ and
\beq
{\cal R}^{\mu\nu}(\theta_R,\phi_R)=
\frac{1}{4M}\Tr\left[\sla{S}_L\overline{J}^\mu
	(\sla{P}_N+M)J^\nu\right]
\label{75pp}
\eeq
which is linear in the bound nucleon spin four--vector $S^\mu _L$.
The angles $\theta_R$, $\phi_R$ define the direction of the spin
${\bd s}_L$ in the frame in which the bound nucleon is at rest. The
explicit expression of ${\cal R}^{\mu\nu}$ can be found in appendix C.
Obviously this is not the only possible way to compute 
eq.~(\ref{37pp}) ---we have used eqs.~(\ref{75a},~\ref{75pp}) because
explicit expressions for the ${\cal R}^{\mu\nu}$'s with the current
operators that we consider here were already available.

\subsection*{2.4 Comparison of RPWIA and PWIA}

In PWIA one defines a free electron proton cross section $\sigma^{ep}$
\beq
\sigma^{ep}= \frac{2\alpha^2}{Q^4}\frac{\varepsilon'}{\varepsilon}
	\eta_{\mu\nu}{\cal W}^{\mu\nu} \label{eq2anew} 
\eeq
to write the differential cross section as
\beq
\left(\frac{d^5\sigma}{d\Omega_e d\varepsilon' d\Omega_N}\right)_{PWIA}=
\frac{p_NMM_B}{M_A f_{rec}}\sigma^{ep}N_{b}(p)
	\label{eq1new}
\eeq
with $N_b(p)$ the momentum distribution of the non--relativistic bound
orbital (eq.~(\ref{10})) normalized to 1 ($\int d{\bd p} N_b(p) =1$).

In analogy, in RPWIA we may define single--nucleon cross sections 
$\sigma^{ep}_{uu}$, $\sigma^{ep}_{vv}$ and $\sigma^{ep}_{uv}$
corresponding to the single--nucleon tensors ${\cal W}^{\mu\nu}$,
${\cal Z}^{\mu\nu}$ and ${\cal N}^{\mu\nu}$ appearing in the
$W^{\mu\nu}_P$, $W^{\mu\nu}_N$ and $W^{\mu\nu}_C$  hadronic tensors:
\beqa
\sigma^{ep}_{uu}&=& \frac{2\alpha^2}{Q^4}\frac{\varepsilon'}{\varepsilon}
	\eta_{\mu\nu}{\cal W}^{\mu\nu} = \sigma^{ep} \label{eq2a} \\
\sigma^{ep}_{vv}&=& \frac{2\alpha^2}{Q^4}\frac{\varepsilon'}{\varepsilon}
	\eta_{\mu\nu}{\cal Z}^{\mu\nu} \label{eq2b} \\
\sigma^{ep}_{uv}&=& \frac{2\alpha^2}{Q^4}\frac{\varepsilon'}{\varepsilon}
	\eta_{\mu\nu}{\cal N}^{\mu\nu} \label{eq2c} \,\,\, .
\eeqa

Using the above definitions together with eqs.~(\ref{e5}),~(\ref{24}) and
(\ref{29pc})--(\ref{63pc}) we can write the differential cross section in
RPWIA (see eq.~(\ref{1})) as
\beq
\frac{d^5\sigma}{d\Omega_e d\varepsilon' d\Omega_N}=
\frac{p_NMM_B}{M_A f_{rec}}\left[
	\sigma^{ep}_{uu}N_{uu}(p)+
	\sigma^{ep}_{vv}N_{vv}(p)+
	\sigma^{ep}_{uv}N_{uv}(p)\right] \,\,\, ,
\label{eq1}
\eeq
where $\sigma^{ep}_{uu}$ is the free electron--proton cross section
appearing in eqs.~(\ref{int1}) and (\ref{eq1new}) 
obtained from positive--energy projections, while 
$\sigma^{ep}_{vv}$ and $\sigma^{ep}_{uv}$ are new components that are
solely due to the negative--energy projections and that may only appear in
scattering from a bound nucleon. Thus, in
RPWIA the differential cross section depends on both positive and
negative energy projections of the relativistic bound nucleon wave function.
Equation~(\ref{eq1}) is self-explanatory: it shows that one can 
perform a decomposition into a part proportional to the square of the
positive--energy projection, $N_{uu}$, with proportionality factor
$\sigma^{ep}_{uu}$. This proportionality factor involves only 
free $u$-Dirac spinors and it is identical to the standard $\sigma^{ep}$
that appears in the non--relativistic treatment (see eq.~(\ref{int1})). The rest of
the cross section is proportional (quadratically and linearly)
to the negative--energy projection ($N_{vv}$ or $N_{uv}$) and the
proportionality factor involves (quadratically or linearly) free $v$-Dirac
spinors. 

From eq.~(\ref{eq1}) it is also a simple matter to go back to the
{\em non--relativistic limit} by imposing condition (\ref{11}) on the bound
nucleon wave function. Indeed if we impose
\beq
\phi^d=\phi^d_{(0)}=\frac{{\bd \sigma}\cdot{\bd p}}
	{\overline{E}+M}\phi^u
	\label{eq50new}
\eeq
we get
\beqa
\tilde{\alpha}_\kappa &=& \tilde{\alpha}_\kappa ^{(0)} =
	\sqrt{\frac{2M}{\overline{E}+M}}\frac{g_\kappa}{\sqrt{4\pi}} 
		\label{51new} \\
\tilde{\beta}_\kappa &=& \tilde{\beta}_\kappa^{(0)} =0 
		\label{52new}
\eeqa
and therefore all the terms containing negative--energy
projections in eq.~(\ref{eq1}) become zero
\beqa
N_{vv}^{(0)} &=& 0  \\
N_{uv}^{(0)} &=& 0  \,\,\, ,
\label{53new}
\eeqa
while the term depending only on positive--energy projections becomes
\beq
N_{uu}(p) = N_{uu}^{(0)}(p) = \frac{2M}{\overline{E}+M}
	\frac{g^2_\kappa (p)}{4\pi}\,\,\, .
	\label{54new}
\eeq
Thus,  we recover eq.~(\ref{eq1}) with the non--relativistic momentum
distribution $N_b(p)$ replaced by $N_{uu}^{(0)}(p)$ in eq.~(\ref{54new}).
Actually to go to the non--relativistic limit we should neglect terms of the
order $(1-\frac{\overline{E}}{M})$ and therefore we write
\beq
N_{uu}^{n.r.}(p) = \frac{g_\kappa ^2(p)}{4\pi}K \,\,\, ,
\label{55new}
\eeq
where $K$ is a normalization factor taking into account that
$g_\kappa (p)$ is not normalized to 1 while $N_{uu}^{n.r.}(p)$ should be
normalized to 1. Hence
\beq
K^{-2} = \int p^2 dp g^2_\kappa (p) \,\,\, .
	\label{56new}
\eeq

We note that $K \geq 1$ because of the normalization condition of
the relativistic bound nucleon wave function (see appendix A).
Therefore in the non--relativistic limit the momentum distribution
$N_{uu}^{n.r.}(p) \geq g_\kappa^2 (p)/4\pi$.
Actually since $f_\kappa^2(p)$ is small and does not contribute by more
than a $3\%$ to the normalization (for the case considered
here $K = 1.025$) one has that  $N_{uu}^{n.r.}(p) \approx N_{uu}^{(0)}(p)$
in the whole momentum range considered. 

In Figure~1 we show the functions $N_{uu}$, $N_{uv}$ and $N_{vv}$
 (in fm$^3$) corresponding to the shell $1p_{1/2}$
in $^{16}$O.
The bound state wave function for the proton has been computed within
the framework of the Walecka relativistic model. The
mean field in the Dirac equation is determined through a Hartree procedure
from a phenomenological relativistic Lagrangian with scalar and vector
(S--V) terms. We use the parameters of ref.~\cite{hor},
and the TIMORA code~\cite{ho91} to get the radial functions
$g_{\kappa_b}$, $f_{\kappa_b}$ in coordinate space and then transform to
momentum space according to the equations in appendix A.

In the left panel of fig.~1 we show the results
for the three projection components of the momentum distribution
$N_{uu}(p)$ (solid), $N_{uv}(p)$ (dotted)
and $N_{vv}(p)$ (dashed) as given by eqs.~(\ref{77p1}--\ref{77p3}).
Note that the components $N_{uu}$ and $N_{vv}$ are positive
whereas $N_{uv}$ can be positive or negative depending on the $p$--value.
This is specified by the signs $(+)/(-)$.
The component $N_{uu}(p)$ clearly dominates in the region
$p \leq 300$ MeV/c. In fact, in the maximum ($p\sim 100$ MeV)
it is one order of magnitude larger than
the $N_{uv}$ component and more than two orders of magnitude larger
than the $N_{vv}$ component.
This suggests that in this region one can expect negligible 
contributions from the components $N_{uv}(p)$ and $N_{vv}(p)$.
Therefore, for low $p$--values, one may expect that
projecting the bound nucleon wave function over positive
energy states gives basically the same result as a
fully relativistic calculation.
This assesment of course needs some care because in order to get final
results one also needs to evaluate the single--nucleon components
$\sigma^{ep}_{uu}$, $\sigma^{ep}_{uv}$ and $\sigma^{ep}_{vv}$.

In the region of high missing momentum, $p > 300$ MeV/c, the 
situation is clearly different.
In this $p$--region the values of the
components $N_{uv}(p)$ and $N_{vv}(p)$ are similar to or even larger than
that of $N_{uu}(p)$. Therefore, for $p>300$ MeV/c one may expect that the
effects of the dynamical enhancement
of the lower components in the bound relativistic wave function
will be observable.

It is also interesting to compare the positive--energy projection of the
full relativistic bound wave function with the non--relativistic limits defined
in eqs.~(\ref{54new},\ref{55new}).
In the right panel of fig.~1 we compare the positive--energy projection
$N_{uu}(p)$ (solid) with $N_{uu}^{(0)}(p)$ (eq.~\ref{54new})
(dotted line) and with
$N_{uu}^{n.r.}(p)$ (eq.~\ref{55new}) (dashed line). 
Note that the difference
between $N_{uu}^{(0)}(p)$ and $N_{uu}^{n.r.}(p)$ is negligible for all
$p$--values. Moreover, the difference between these two functions and
the component $N_{uu}(p)$ is only visible 
for $p \geq 400$ MeV/c. Since up to $p\simeq 400$ MeV there are no
important differences between $N_{uu}$ and $N_{uu}^{n.r.}$, we shall speak
roughly of positive--energy projected and non--relativistic limits when
discussing total nuclear responses and differential cross sections in
section 4.

\subsection*{2.5 Longitudinal-Transverse separation}

To finish with this section we note that one
may also use the decomposition of currents into 
longitudinal and tranverse components and then write the different
projections of the $\sigma$'s as follows
\beqa
\sigma^{ep}_{uu} &=& \sigma_{MOTT}\left(\sum_K v_K {\cal R}^K_{uu} \right)
\label{eq3a} \\
\sigma^{ep}_{vv} &=& \sigma_{MOTT}\left(\sum_K v_K {\cal R}^K_{vv} \right)
\label{eq3b} \\
\sigma^{ep}_{uv} &=& \sigma_{MOTT}\left(\sum_K v_K {\cal R}^K_{uv} \right)
\label{eq3c} 
\eeqa
with $K=L,T,TL,TT$  the longitudinal, transverse,
TL interference and extra tranverse contributions of
the hadronic tensor for free nucleons. 
The single--nucleon responses ${\cal R}^K_{uu}$,
${\cal R}^K_{vv}$ and ${\cal R}^K_{uv}$ are
given by taking the appropiate components of the single--nucleon tensors
${\cal W}^{\mu\nu}$, ${\cal Z}^{\mu\nu}$ and ${\cal N}^{\mu\nu}$,
as is done for the $R^K_{uu}$ in PWIA. It should
be pointed out that contrary to the situation that occurs for on--shell
nucleons, different results may be obtained depending on the choice of the
current operator. In particular they depend on whether current
conservation is or not imposed to eliminate one of the components, and on
which of the 0- or 3- components is eliminated.
This subject was discussed in detail in refs.~\cite{dp,fo,naus}
for the case of the ${\cal R}^K_{uu}$ responses with and without
polarizations and will be discussed in the next section for the
${\cal R}^K_{uv}$ and ${\cal R}^K_{vv}$ as well. Obviously,
the hadronic response functions can be also written in RPWIA in the form
\beq
R^K=R^K_{P}+R^K_{N}+R^K_{C} 
\label{eqx}
\eeq
with each component given by
\beqa
R^K_{P}&=&{\cal R}^K_{uu}N_{uu}(p) \label{eq5a} \\
R^K_{N}&=&{\cal R}^K_{vv}N_{vv}(p) \label{eq5b} \\
R^K_{C}&=&{\cal R}^K_{uv}N_{uv}(p) \label{eq5c} \,\,\, .
\eeqa

In standard PWIA only the responses ${\cal R}^K_{uu}$ 
which are due solely to the 
positive--energy components occur. In this case, by dividing the experimental
5-differential cross section by $\sigma^{ep}_{uu}$ and the 
kinematical factors, one could measure directly the momentum
distribution $N_{uu}(p)$.
However, in the more general case of
RPWIA, as we see in eqs.~(\ref{eq1},\ref{eqx}), this is not possible in
general, as we have additional contributions from the 
negative--energy components that, {\em a priori}, cannot be isolated
by  experimental procedures. One could, however, study under what
circumstances the results of eqs.~(\ref{eq1},\ref{eqx}) are dominated
by some of the $uu$, $vv$ or $uv$ contributions.
We present in detail this study in section 4 where results
for the various observables of the scattering reaction
are shown.

Analogously to the analysis within the standard PWIA~\cite{dp,fo,naus},
the evaluation of
the three projection components of the various single--nucleon quantities
depends on what single--nucleon current operator is chosen,
on whether current conservation is fulfilled or imposed, and in the latter
case on how it is imposed.
This study is presented in detail in section
3, where different
prescriptions involving single--nucleon current operators and the
continuity equation are considered. Here we simply mention that there exist
some kinematics for which the $uu$, $vv$ or $uv$ contributions to the
single--nucleon
cross section and single--nucleon response functions 
are comparable, and therefore a fully
relativistic analysis
can give very different results from the standard PWIA.

\section*{3. Current operators and single--nucleon responses}

\subsection*{3.1 Choices of the current operator for bound nucleons}

As discussed in several works~\cite{fm85,dp,fo,naus} the choice of the current
operator $\hat{J}^\mu_N=\hat{J}^\mu$ is to some extent arbitrary. 
We discuss here the most popular
choices denoted as CC1 and CC2 in ref.~\cite{fo}.

The current operator CC1
\beq
\hat{J}^\mu_{CC1}=(F_1+F_2)\gamma^\mu-
\frac{F_2}{2M}\left(\overline{P}+P_N\right)^\mu
\label{c5}
\eeq
with $\overline{P}^\mu=(\overline{E},{\bd p})$ 
($\overline{E}=\sqrt{p^2+M^2}$ and ${\bd p}={\bd p}_N-{\bd q}$)
is obtained from the current operator CC2
\beq
\hat{J}^\mu_{CC2}= F_1\gamma^\mu +i\frac{F_2}{2M}\sigma^{\mu\nu}Q_\nu
\label{c4}
\eeq
replacing $Q_\nu$ by $(P_N-\overline{P})_\nu$ and
using the Gordon decomposition for the free nucleon case, i.e., assuming
that initial and final nucleons in eq.~(\ref{e3b}) satisfy the free Dirac equation
$(\sla{P}-M)\Psi =0$. For free nucleons eqs.~(\ref{c5}) 
and (\ref{c4}) are totally
equivalent and, since the current is conserved, the time component can be
eliminated by writing it in terms of the third component or vice versa.
In the case discussed here the final nucleon in eq.~(\ref{e3b}) satisfies the free
Dirac equation, but the initial bound nucleon does not, and therefore the
nucleon current and the nuclear responses are different when using CC1 or
CC2 operators. Moreover the current is not conserved and, once the current
operator (CC1 or CC2) has been chosen, one may choose to impose current
conservation by eliminating the third component (CC1$^{(0)}$or
CC2$^{(0)}$), or the time component (CC1$^{(3)}$ or CC2$^{(3)}$). One may
also choose not to impose current conservation (NCC1 or NCC2). In general
each of these six different choices gives a different result and it is not
clear how to define criteria that favour one choice over the others (see
for instance~\cite{naus}). In the absence of sharper criteria it is advisable to
use as reference a current operator that guarantees current conservation
for any initial and final nucleon wave function. For this purpose we
choose the operator
\beq
\hat{J}^\mu_{C}=F_1\gamma^\mu+i\frac{F_2}{2M}\sigma^{\mu\nu}Q_\nu
-F_1\frac{Q^\mu \sla{Q}}{Q^2}
\label{c3}
\eeq
for $Q^2 <0$, which is also equivalent to the CC1 and CC2 operators in the
free nucleon case. Although we define the operator $\hat{J}^\mu_C$ only for
the case of virtual photon ($Q^2< 0$), we note that the last term in
eq.~(\ref{c3}) has no transverse component and therefore it does not contribute
in the case of real photons anyway, where one could question its validity
because of the divergence as $Q^2 \rightarrow 0$. This operator has
several good features: a) with this operator
the current in eq.~(\ref{e3b}) is conserved, and b) the
amplitude ${\cal M}$ in eq.~(\ref{e1}) is invariant under the replacement of
$\hat{J}^\mu_C$ by $\hat{J}^\mu_{CC2}$ because the electron current is
conserved, which implies that the last term in eq.~(\ref{c3}) does not contribute
to the amplitude ${\cal M}$. A similar operator was introduced by 
Gross and Riska in ref.~\cite{riska}.

\subsection*{3.2 Comparison of different currents and single--nucleon responses}

To  understand better the results shown in the next
sections it is helpful to split the amplitude in eq.~(\ref{e1}) into its
longitudinal (L) and transverse (T) components
\beq
{\cal M}={\cal M}^L-{\cal M}^T
\label{eq59}
\eeq
with
\beqa
{\cal M}^L&=& = j_e^0\left(J^0-\frac{w}{q}J^3\right)
\label{c6a} \\
{\cal M}^T&=&{\bd j}_e^T {\bd J}^T = j_e^x J^x+j_e^y J^y \,\,\, ,
\label{c6}
\eeqa
where we have taken the $z$--axis parallel to ${\bd q}$, and used the
continuity equation for the electron current 
($\omega j_e^0 = {\bd q}\cdot{\bd j_e}$).
Now it is simple to check that the hadronic longitudinal current 
that we define for any current operator as
\beq
J^L=(J^0-\frac{\omega}{q}J^3)
\label{a1}
\eeq
is identical for
the $J^\mu_{C}$ and $J^\mu_{CC2}$ choices. Indeed
\beq
J^L_{C}=\left(J^0_{C}-\frac{w}{q}J^3_{C}\right)=
-\frac{Q^2}{q^2}
J^0_{C}=\left(J^0_{CC2}-\frac{w}{q}J^3_{CC2}\right) =
J^L_{CC2} \,\,\, .
\label{c7}
\eeq
On the other hand since the transverse components are equal,
\beq
{\bd J}^T_{CC2}={\bd J}^T_{C}\,\,\, ,
\label{c8}
\eeq
both the longitudinal
${\cal M}^L$, and the transverse ${\cal M}^T$ amplitudes (and hence each of the response
functions) are invariant separately under the replacement of $\hat{J}_{C}$ 
by $\hat{J}_{CC2}$. We emphasize that
this is so provided that the complete expression for $J^L_{CC2}$ 
in eq.~(\ref{c7}) is
used. We note however that if one starts from the CC2 operator and
imposes current conservation (as done in the CC2$^{(0)}$ and CC2$^{(3)}$
versions), the longitudinal amplitude changes. In the CC2$^{(0)}$ version,
one replaces $J^3_{CC2}$ by $\displaystyle\frac{\w}{q}J^0_{CC2}$ and therefore
$J^L_{CC2}$ is replaced by
\beq
J^L_{CC2}\rightarrow J^L_{CC2(0)}=\frac{-Q^2}{q^2}
J^0_{CC2} \,\,\, .
\label{c9}
\eeq
Similarly in the CC2$^{(3)}$ version one replaces $J^0_{CC2}$ by 
$\displaystyle\frac{q}{\w}J^3_{CC2}$, or equivalently $J^L_{CC2}$ is
replaced by 
\beq
J^L_{CC2}\rightarrow J^L_{CC2(3)}=\frac{q}{\w}
	\left(\frac{-Q^2}{q^2}\right)J^3_{CC2} \,\,\, .
\label{c10}
\eeq
In summary the NCC2 choice is equivalent to using the conserved current defined
in eq.~(\ref{c3}). To the contrary, with
the CC2$^{(0)}$ or CC2$^{(3)}$ choices the 
${\cal M}^L$ amplitude becomes different and so do the response functions $R^L$
and $R^{TL}$, while the transverse amplitude and response functions remain
unchanged.

The effect of going from the conserved current to the 
CC2$^{(0)}$ or CC2$^{(3)}$ choices can be easily evaluated by considering
the differences
\beq
\Delta^L_{CC2(0)}\equiv J^L_{CC2}-J^L_{CC2(0)}
\label{c11}
\eeq
and 
\beq
\Delta^L_{CC2(3)}\equiv J^L_{CC2}-J^L_{CC2(3)}
\label{c12} \,\,\, .
\eeq
As expected they are both proportional to $\Psi_N\sla{Q}\Psi_i$, with
different proportionality factors,
\beqa
\Delta^L_{CC2(0)}&=&\frac{\w^2}{q^2}F_1\Psi_N \frac{\sla{Q}}{\w}\Psi_i
\label{c13} \\
\Delta^L_{CC2(3)}&=&F_1\Psi_N\frac{\sla{Q}}{\w}\Psi_i \,\,\, .
\label{c14}
\eeqa

Similar relationships hold when 
the CC1$^{(0)}$ and CC1$^{(3)}$ currents are compared to NCC1, but in this
case the differences $\Delta^L_{CC1(0)}$ and $\Delta^L_{CC1(3)}$ also contain
a term proportional to $(1-\overline{\omega}/\omega)$.
Indeed when comparing $J^L_{CC1(0)}$ and $J^L_{CC1(3)}$ to $J^L_{CC1}$ one
gets
\beqa
\Delta^L_{CC1(0)}&\equiv & J^L_{CC1}-J^L_{CC1(0)}=
\overline{\Psi}_N\frac{\w^2}{q^2}\left[(F_1+F_2)\frac{\sla{Q}}{\w}-
	\frac{F_2(E_N+\overline{E})}{2M}\left(1-\frac{\overline{\w}}{\w}
	\right)\right]\Psi_i
\nonumber \\
&  &
\label{c18a} \\
\Delta^L_{CC1(3)}&\equiv & J^L_{CC1}-J^L_{CC1(3)}=
\overline{\Psi}_N
\left[(F_1+F_2)\frac{\sla{Q}}{\w}-
	\frac{F_2(E_N+\overline{E})}{2M}\left(1-\frac{\overline{\w}}{\w}
	\right)\right]\Psi_i \,\,\, .
\nonumber \\
\label{c18b}
\eeqa
Obviously the transverse currents satisfy 
\beqa
& & J^T_{CC1(0)}=J^T_{CC1(3)}=J^T_{CC1} \nonumber \\
& & J^T_{CC2(0)}=J^T_{CC2(3)}=J^T_{CC2} \,\,\, . \nonumber 
\eeqa

Finally we compare NCC2 to NCC1. We note that $\hat{J}^\mu_{CC2}$ can be
written without lack of generality as
\beqa
\hat{J}^\mu_{CC1}-\hat{J}^\mu_{CC2}&=& 
\frac{F_2}{2M}\left[\left\{(M-\sla{P}_N)
\gamma^\mu+\gamma^\mu(M-\sla{\overline{P}})\right\}
+(E+E_N-\overline{E}-\overline{E}_N)\delta_{\mu,0} \right]
\nonumber \\
&=&
\frac{F_2}{2M}\left[\left\{(M-\sla{P}_N) 
\gamma^\mu+\gamma^\mu(M-\sla{\overline{P}})\right\}
-\delta_{\mu,i}(\w-\overline{\w})\gamma^0 \gamma^i \right]
\,\, ,i=1,2,3 \,\, . \nonumber \\
& &  \label{c21b}
\eeqa

From eq.~(\ref{c21b}) it is easy to see that all of the components are different for
the NCC2 and NCC1 choices when $\Psi_i$ is not a free $u$--spinor
\beq
J^\mu_{CC2}\neq J^\mu_{CC1}, \,\,\, \mu=0,1,2,3
\label{eq74}
\eeq
and
\beqa
J^L_{CC2} &\neq & J^L_{CC1} 
\label{eq75}
\\
J^T_{CC2} &\neq & J^T_{CC1} \,\,\, .
\label{eq76}
\eeqa
We shall discuss these differences in more detail in the next subsection.
However it is also interesting to compare the positive--energy projections
of $J^\mu_{CC2}$ and $J^\mu_{CC1}$ (the $u$--parts defined in section 2).
For the positive--energy projections one sees that
\beq
\langle J^0_{CC2} \rangle_{u}=\langle J^0_{CC1} \rangle_{u}
\label{eq77}
\eeq
and
\beq
\langle J^i_{CC2}\rangle_{uu} =
\langle J^i_{CC1}\rangle_{uu} +
\frac{F_2}{2M}(\w-\overline{\w})
\overline{u}_N\gamma^0\gamma^i u \,\,\,\, ,\,\,\,\,\,\, i=1,2,3 \,\, .
\label{c25}
\eeq
Hence, when considering positive--energy projections one has that the time
components are equal and the transverse and longitudinal components differ
by an amount proportional to $\omega-\overline{\omega}$. This in turn
implies that
\beq
\langle J^L_{CC2(0)}\rangle_{u}=
\langle J^L_{CC1(0)}\rangle_{u} \,\,\, ,
\label{eq79}
\eeq
while
\beq
\langle J^L_{CC2}\rangle_{u}-\langle J^L_{CC1}\rangle_{u}
\propto (\omega-\overline{\omega})/2M
\label{eq80}
\eeq
and
\beq
\langle J^T_{CC2}\rangle_{u}-\langle J^T_{CC1}\rangle_{u}=
\langle J^T_{CC2(0)}\rangle_{u}-\langle J^T_{CC1(0)}\rangle_{u}=
\langle J^T_{CC2(3)}\rangle_{u}-\langle J^T_{CC1(3)}\rangle_{u}
\propto (\omega-\overline{\omega})/2M \,\,\, .
\label{eq81}
\eeq

The above relationships are important in understanding how
the different single--nucleon responses and the single--nucleon cross
sections behave depending on the choice of the current operator and
depending on whether we consider the positive--energy or
negative--energy components. We show the behaviour of the various response
functions for three different kinematical situations in figures 2--4. 
Figs.~2 and 3 are for kinematics I and II, respectively (see section 2). As
seen in figs. 2a to 4a the longitudinal (L) and transverse-longitudinal
(TL) responses tend to change for each of the current choices (NCC1, NCC2,
CC1$^{(0)}$, CC2$^{(0)}$, CC1$^{(3)}$ or CC2$^{(3)}$),
while the transverse responses (T and
TT) only depend on whether the CC1 or the CC2 current operator is used
(figs.~2b to 4b).
This follows from the fact that the transverse current is independent of
whether current conservation is imposed or not. 

Let us first discuss the
transverse responses whose behaviour is somewhat simpler to understand. 
We recall that NCC2 is equivalent to using the conserved current $J^\mu_C$
and therefore we use it as a reference. As
seen in fig.~2b when CC2 (thick line) is replaced by CC1 (thin line)
the positive--energy component
of the transverse responses ${\cal R}^{T,TT}_{uu}$ changes little
while the negative--energy components 
${\cal R}^{T,TT}_{uv}$ and ${\cal R}^{T,TT}_{vv}$ differ by an order of
magnitude or more. The reason for this is easily understood by looking at 
eq.~(\ref{c21b}). When considering the positive--energy component the first term
in this equation becomes zero ($(\sla{\overline{P}}-M)u=0)$ 
and the difference between
the two curves for ${\cal R}^{T}_{uu}$ is proportional to $(\omega-
\overline{\omega})/2M$ (see eqs.~(\ref{c25}),~(\ref{eq81})); 
moreover in the interference
response function ${\cal R}^{TT}_{uu}$ the $(\omega-
\overline{\omega})/2M$ dependence tends to cancel. On the contrary when
one considers the negative--energy components
${\cal R}^{T,TT}_{uv}$ and ${\cal R}^{T,TT}_{vv}$ the first term in
eq.~(\ref{c21b}) is maximum ($(\sla{\overline{P}}+M)v=2M$) and the
difference between CC1 and CC2 is maximized. 
Although quantitative details depend on the kinematics, the same
qualitative behaviour is seen in figs.~3b and 4b. It is also interesting
to remark that with the CC2 choice the negative--energy components are of
the same order as the positive--energy ones in the whole $p$--region
considered, while with the CC1 current 
(because of $(M-\sla{\overline{P}})$ terms) the
$uv$ and $vv$ components may become much larger than the $uu$ ones.
Hence one can expect a stronger dependence on the negative--energy
projections $\beta_\kappa$ when using the CC1 operator.

These trends are also observed in the longitudinal and
longitudinal--transverse responses (figs. 2a to 4a). The CC1 current gives
similar responses as the CC2 current for the $uu$ components, but for the
$uv$ and $vv$ components, the CC1 current tends to give much larger results
(for the absolute values). In addition, a very important role is played by
the way in which current conservation is imposed (or not). 
If one chooses not to
impose current conservation (NCC1 or NCC2) one observes the same
qualitative behaviour as that just discussed for the transverse
components. The choice of 
imposing current conservation by elimination of the third component 
(CC1$^{(0)}$ and CC2$^{(0)}$) also shows a behaviour similar to that
discussed above 
and the results tend to be close to the NCC1, NCC2 responses.
Particularly in this case CC1$^{(0)}$ and CC2$^{(0)}$ give identical results
for the $uu$ component (as is clear from eq.~(\ref{eq79})).
However the CC1$^{(3)}$ and CC2$^{(3)}$ choices
cause large deviations in all of the components including the positive--energy
components ${\cal R}^{L,TL}_{uu}$, particularly for kinematics II.
This behaviour is mainly due to the $\Psi_i
\sla{Q}\Psi_f$ term in eqs.(\ref{c14},\ref{c18b}) that 
(as seen in figs. 2a to 4a) may cause
a strong reduction of the ${\cal R}^{L,TL}_{uu}$ responses when one goes
from NCC1 to CC1$^{(3)}$ or from NCC2 to CC2$^{(3)}$.
This reduction is also present
in the CC1$^{(0)}$ and CC2$^{(0)}$ choices, but in these cases
it is largely attenuated by the 
$\omega^2/q^2$ factor in eqs.~(\ref{c13},\ref{c18a}) and therefore the
latter choices give results much closer to the NCC2 result that we use as
reference. 

It is important to remark that the simple relations stated in
eqs.~(\ref{c13}--\ref{c18b}) are also responsible
for the large differences observed between results
obtained with CC1$^{(0)}$ and CC1$^{(3)}$ choices (or with 
CC2$^{(0)}$ and CC2$^{(3)}$)  in the context of PWIA
(see refs.~\cite{dp,fo,naus}). In these references the study was
restricted to $\sigma^{ep}$ (or its various polarization components), although
the cause of the large differences was not as clearly identified as it is
here. In 
particular, the following interesting relationship between TL responses
follows: 
\beq
\frac{{\cal R}^{TL}_{CCi}-{\cal R}^{TL}_{CCi(0)}}
{{\cal R}^{TL}_{CCi}-{\cal R}^{TL}_{CCi(3)}}= \frac{\omega^2}{q^2}
\ll 1, \,\,\,\,\,\,\,\, i=1,2 \,\, .
\label{83new}
\eeq
A more involved relationship can be established for the ${\cal R}^L$
responses obtained with different choices.

The single--nucleon cross section components $\sigma^{ep}_{uu}$, 
$\sigma^{ep}_{uv}$
and $\sigma^{ep}_{vv}$ are also shown for completeness in figs.~5 and 6. The
results shown correspond to a redefinition of the $\sigma$'s including the
factor $M^2/\overline{E} E_N$ for easy comparison to previous work along
similar lines, i.e.,
\beq
\tilde{\sigma}^{ep}_{uu}=\frac{M^2}{\overline{E}E_N}\sigma^{ep}_{uu}
=\sigma^{ep}
\label{eq82}
\eeq
with $\sigma^{ep}$ as defined by de Forest~\cite{fo}. Similar definitions
are used for $\tilde{\sigma}^{ep}_{uv}$ and $\tilde{\sigma}^{ep}_{vv}$ 
in terms of $\sigma^{ep}_{uv}$ and $\sigma^{ep}_{vv}$, respectively.

Fig.~5 corresponds to kinematics I for $\theta_e=30^0$ (top panel) and
$\theta_e=150^0$ (lower panel) and fig.~6 corresponds to kinematics II with
$\theta_e=12.5^0$ (the forward-angle limit of the Hall
A spectrometers at CEBAF).
For the positive--energy components ($\sigma^{ep}_{uu}$)
the dependence on the different choices of the current operator
is maximal for kinematics I and forward angle.
For kinematics II that dependence remains small at any angle.
The dependence of $\tilde{\sigma}^{ep}_{uu}$ (i.e.,
$\sigma^{ep}$) on the current operator
has been discussed at length
in refs.~\cite{dp,fo,naus}. Much more notorious is the dependence of
$\sigma^{ep}_{uv}$ and $\sigma^{ep}_{vv}$ on the various current 
choices. 
Obviously the behaviour of the single--nucleon cross sections at 30$^0$ in
fig.~5 mainly reflects the behaviours of ${\cal R}^L$ (and partly of
${\cal R}^{TL}$), while the behaviour of $\sigma$'s at $\theta_e=150^0$
(bottom in fig.~5) mainly results from the interplay between 
${\cal R}^{TL}$ and ${\cal R}^T$ responses. Compared to the wild variation
of $\sigma^{ep}_{uv}$ and $\sigma^{ep}_{vv}$, $\sigma^{ep}_{uu}$ shows a
rather mild dependence on the choice of current operator.

We would like
to emphasize that the changes in $\sigma^{ep}_{uu}$
are minor for NCC1, NCC2, CC1$^{(0)}$ 
and CC2$^{(0)}$ and are important for CC1$^{(3)}$ and CC2$^{(3)}$,
particularly at small $q$ (kinematics I). As one increases
$q$ ($q\geq 1$ GeV) the latter choices may not be that poorly behaved.

It is also important to remark here that
the single--nucleon cross sections associated with 
the negative--energy projections
($\sigma^{ep}_{uv}$ and $\sigma^{ep}_{vv}$) tend to be larger in absolute
values when the CC1 current operator is used. Therefore one may expect a larger
effect of the negative--energy projections in the total differential cross 
sections when using $\hat{J}^\mu_{CC1}$, and therefore a larger
deviation from the non--relativistic PWIA result.

\subsection*{3.3 The effect of S--V potentials}

We now study in detail the problem of different current operator
choices for nucleons
satisfying the Dirac equation with S--V potentials
\beq
(\sla{P}-M-V\gamma_0 +S)\Psi=0\,\,\, .
\label{c30}
\eeq
In principle both initial and final nucleons could be considered as
solutions of this equation with the same potential;
however, in the present work to simplify the problem somewhat in
discussing the situation where the final nucleon is at relatively high
energies (and so roughly quasifree)
here we concentrate on the case (RPWIA) in which the
outgoing nucleon is a solution of the free Dirac equation.
Thus for the final wave function we have
\beq
(\sla{P}-M)\Psi_f=0 \,\,\, ,
\label{c31}
\eeq
while for the initial
\beq
(\sla{P}-M)\Psi_i=(V\gamma_0 -S)\Psi_i \,\,\, .
\label{c32}
\eeq

To understand the effect of the S--V potentials in the initial state
we first write the wave function in coordinate space
\beq
\Phi(x_\mu )=e^{-iEt}\Psi({\bd x})=e^{-iEt}
		\left(\begin{array}{@{\hspace{0pt}}c@{\hspace{0pt}}}
		\phi_{u}({\bd x}) \\
		\phi_{d}({\bd x})\end{array}\right)
\label{c33}
\eeq
with
\beq
\phi_d({\bd x})=\frac{{\bd \sigma}\cdot{\bd p}\phi_u({\bd x})}
	{\tilde{E}({\bd x})+\tilde{M}({\bd x})} \,\,\, ,
\label{c34}
\eeq
where ${\bd p} = -i{\bd \nabla}$ and
\beqa
\tilde{E}({\bd x})&=&E-V(x)=\tilde{E}(x) \nonumber \\
\tilde{M}({\bd x})&=&M-S(x)=\tilde{M}(x)\,\,\, .
\label{c35}
\eeqa
For $S$ and $V$ constant (nuclear matter) the solutions are still plane waves
satisfying 
\beq
\tilde{E}_0=\sqrt{{\bd k}^2+\tilde{M}_0^2}\,\,\,\,\,\,\,\,
(k\leq k_F)\,\,\,\, ,\,\,\,\,\, \tilde{E}_0=E-V_0
\,\,\,\, ,\,\,\,\,\, \tilde{M}_0= M-S_0 \,\,\, ,
\label{c36}
\eeq
while for $S$ and $V$ $x$--dependent the solutions satisfy the standard
equations
\beqa
\nabla^2\phi_{u}({\bd x})&=&-(\tilde{E}^2-\tilde{M}^2)\phi_u-
i\eta'(x){\bd \sigma}\cdot {\bd x}\phi_d \nonumber \\
\nabla^2\phi_{d}({\bd x})&=&-(\tilde{E}^2-\tilde{M}^2)\phi_d+
i\eta(x){\bd \sigma}\cdot {\bd x}\phi_u 
\label{c37}
\eeqa
with
\beqa
\eta'&=&\frac{1}{x}\frac{d}{dx}(S+V) \nonumber \\
\eta &=& \frac{1}{x}\frac{d}{dx}(S-V) \,\,\, .
\label{c38}
\eeqa

We are interested in working in momentum space and thus write (see also
appendix A)
\beq
\phi_u({\bd p})=\frac{1}{(2\pi)^{3/2}}\int d{\bd x}e^{-i{\bd p}
\cdot {\bd x}}\phi_u({\bd x})
\label{c39}
\eeq 

\beq
\phi_d({\bd p})=\frac{1}{(2\pi)^{3/2}}\int d{\bd x}e^{-i{\bd p}
\cdot {\bd x}}\phi_d({\bd x}) 
=\frac{1}{(2\pi)^{3/2}}\int d{\bd x}e^{-i{\bd p}\cdot {\bd x}}
\frac{(-i{\bd \sigma}\cdot{\bd \nabla}\phi_u({\bd x}))}
{\tilde{E}+\tilde{M}} 
\,\,\, .
\label{c40}
\eeq

We now write (either in ${\bd p}$ or ${\bd r}$--space)\footnote{In this
section ${\bd p}$ is used both for operator (when writing in ${\bd
r}$--space) and as c--number (when writing in momentum space)}
\beq
\Psi=\left[\left(\begin{array}{@{\hspace{0pt}}c@{\hspace{0pt}}}
		1 \\
		\frac{{\bd \sigma}\cdot {\bd p}}
		{\overline{E}+M}\end{array}\right)
	\phi_u+\delta \Psi \right]
\label{c43}
\eeq
with
\beq
\delta\Psi=\left(\begin{array}{@{\hspace{0pt}}c@{\hspace{0pt}}}
		0\\
		\phi'_{d}\end{array}\right)
\label{c43a}
\eeq
and
\beq
\phi'_d=\phi_d-\frac{{\bd \sigma}\cdot{\bd p}}{\overline{E}+M}\phi_u
\,\,\, .
\label{c44}
\eeq

The first term in eq.~(\ref{c43}) is proportional to the positive--energy
projection of $\Psi$ (it has zero negative--energy projection),
while the
second term is a correction proportional to the negative--energy
projection of $\Psi$.
We use this decomposition to analyze the dependence on the
dynamical enhancement ($\phi'_d$) of the bound nucleon wave function when
different choices of the current operator are used. 

We start as before
from CC2 and CC1 and particularize to the present initial and final
nucleon wave functions satisfying eqs.~({\ref{c32}) and (\ref{c31}),
respectively. Using eq.~(\ref{c21b}) together with 
eqs.~(\ref{c31}) and (\ref{c32}) we find that 
\beq
J^\mu_{CC2}-J^\mu_{CC1}=\frac{F_2}{2M}\overline{\Psi}_f\left[
\gamma^\mu(V\gamma^0-S)+(\omega-\overline{\omega})\delta_{\mu,0}\right]
\Psi_i
\label{c46}
\eeq
with 
$\omega-\overline{\omega}=\overline{E}-E$, $\overline{E}=
\sqrt{p^2+M^2}$
and $E$ the energy of the bound nucleon wave function ($E=E_N-\omega=
M_A-M_B^\ast$, neglecting the recoil energy of the residual nucleus).
Hence if we write down explicitly the $\mu =0,1,2,3$ components we find that
\beqa
J^0_{CC2}-J^0_{CC1} &=& \frac{F_2}{2M}\overline{\Psi}_f
[V-S\gamma^0+(\omega-\overline{\omega})]\Psi_i 
\label{c47}\\
J^3_{CC2}-J^3_{CC1} &=& \frac{F_2}{2M}\overline{\Psi}_f
[\gamma^3 (V\gamma^0-S)]\Psi_i
\label{c48} \\
{\bd J}^\perp_{CC2}-{\bd J}^\perp_{CC1} &=& \frac{F_2}{2M}\overline{\Psi}_f
[{\bd \gamma}^\perp(V\gamma^0-S)]\Psi_i \,\,\, .
\label{c49}
\eeqa
For the longitudinal components that enter in NCC2 and NCC1 one gets:
\beqa
J^L_{CC2}-J^L_{CC1}&=&\frac{F_2}{2M}\overline{\Psi}_f
\left[V-S\gamma^0+(\omega-\overline{\omega})-\frac{\omega}{q}
\gamma^3(V\gamma^0-S)\right]\Psi_i 
\nonumber \\
&=& \frac{F_2}{2M}\overline{\Psi}_f
\left(\begin{array}{@{\hspace{0pt}}c@{\hspace{0pt}}}
	[V-S+\omega-\overline{\omega}]\phi_u	
	+\displaystyle\frac{\omega}{q^2}{\bd \sigma}\cdot{\bd q}(V+S)\phi_d
	\\
	\displaystyle\frac{\omega}{q^2}{\bd \sigma}\cdot {\bd q} (V-S)\phi_u
	+[V+S+\omega-\overline{\omega}]\phi_d
	\end{array}\right)
\label{c50}
\eeqa

Hence there are in general three types of terms in the difference between
the longitudinal CC2 and CC1 currents:
\begin{itemize}
\item One proportional to $\omega-\overline{\omega}$ that is present even
after projection into the positive--energy sector. We recall that only in 
CC1$^{(0)}$ and CC2$^{(0)}$ choices the longitudinal currents are equal
for the positive--energy projections (see eq.~(\ref{eq80})).
\item One proportional to the {\sc small} nuclear ($V-S$) potential acting
on the upper component $\phi_u$.
\item One proportional to the {\sc large} nuclear potential ($S+V)$ acting
on the down component $\phi_d$.
\end{itemize}

The latter contribution tends to enhance the role of the negative--energy
components. In other words any dynamical enhancement $\phi'_d$ of the lower
component of the bound nucleon wave function may appear unphysically
augmented in the nuclear response due to lack of current conservation.
Similar considerations apply to the difference between the transverse CC2
and CC1 currents,
\beq
J^x_{CC2}-J^x_{CC1}=\frac{F_2}{2M}
\left(\begin{array}{@{\hspace{0pt}}c@{\hspace{0pt}}}
		-\sigma^x (V+S)\phi_d\\
		\sigma^x(V-S)\phi_u
		\end{array}\right)
\label{c51}
\eeq

\section*{4. Dependence of the response functions and reduced cross sections
on the negative energy projections}

In this section we present results of response functions and differential
cross sections for $1p_{1/2}$ proton knock--out from $^{16}$O. The aim
here is to make a quantitative analysis of the relative importance of
negative--energy projection contributions for various choices of the
current operator. 

The total hadronic response functions in RPWIA are shown in fig.~7. We
have chosen kinematics I and the forms of current operators discussed in
section 3. As already mentioned in that section the transverse responses 
($R^T$ and $R^{TT}$) depend only on whether the current operator CC1 or
CC2 is chosen, while the $R^L$ and $R^{TL}$ responses depend also on the
current conservation prescription. We recall that NCC2 is equivalent to
using an exactly conserved current.

As expected the longitudinal response $R^L$ is practically identical for
NCC2, CC1$^{(0)}$ and CC2$^{(0)}$ (also for NCC1, not shown), while it is
substantially  different for CC2$^{(3)}$. It is interesting to note that,
unlike in PWIA, in RPWIA the longitudinal response $R^L$ for CC1$^{(3)}$ is
closer to the CC1$^{(0)}$ than to the CC2$^{(3)}$ cases. The shift upwards
of $R^L$ in the CC1$^{(3)}$ case is due to the negative--energy projection
$R^L_N$ because, as seen in fig.~2a, the ${\cal R}^L_{vv}$ single--nucleon
response is much larger than ${\cal R}^L_{uu}$, compensating for the
smallness of the negative--energy projection of the wave function 
($N_{vv}$) (see fig.~1). Clearly if we were to consider only 
positive--energy projections, similar
results would be obtained with CC1$^{(3)}$ and
CC2$^{(3)}$ (as is the case in PWIA) ---the term ${\cal R}^L_{vv}$ is
roughly an order of magnitude larger with CC1$^{(3)}$ than with 
CC2$^{(3)}$ causing a larger longitudinal response for the relativistic
bound nucleon. In the longitudinal response function $R^L$ the
contributions of the negative--energy projections are maximized for
CC1$^{(3)}$, are minimized for NCC2 and are also negligible for
CC1$^{(0)}$ and CC2$^{(0)}$. This is illustrated in the top panel of
fig.~8 where $R^L_P$, $R^L_N$ and $R^L_C$ are plotted separately for
various current choices. Here we use $P$, $N$ and $C$ to denote positive,
negative and cross terms, respectively.
In the CC1$^{(3)}$ case $R^L_P$ is only $80\%$ of
the total $R^L$, while for the other choices $R^L_P$ is more than $95\%$ of
the total (see also figs.~10,~11).

As seen in fig.~7 the $R^{TL}$ response is the one where the effects of
different prescriptions for the nucleon current are the largest. The
maximum of this response can change by as much as a factor of 2 (or 3)
when one goes from CC1$^{(3)}$ to CC2$^{(3)}$ and CC1$^{(0)}$ (or to
CC2$^{(0)}$). Here one can see with the help of the lower panel of fig.~8
(see also figs.~10, 11) that the negative--energy projections of the bound
nucleon play a very important role through the crossed term $R^{TL}_C$,
while $R^{TL}_N$ remains negligible. This follows from the fact that in
the $p$ region under study, ${\cal R}^{TL}_{uv}$ is of the same order as 
${\cal R}^{TL}_{vv}$ and much larger than ${\cal R}^{TL}_{uu}$ (see
fig.~3a). As in the case of $R^L$ also for $R^{TL}$ the negative--energy
contributions are minimal with the CC2$^{(0)}$ choice and maximal with the
CC1$^{(3)}$ choice. With this latter choice the negative--energy
contribution $R^{TL}_C$ is dominant, contributing by $\sim 75\%$ to the
total $R^{TL}$, with CC1$^{(0)}$ and CC2$^{(3)}$ choices $R^{TL}_C$ and 
$R^{TL}_P$ contributions are approximately equal and even with the 
CC2$^{(0)}$ choice the $R^{TL}_C$ cross term gives a sizable ($\sim 30\%$)
contribution. Since for all the choices the positive--energy projections in
the maximum are similar, the net $R^{TL}$ responses in RPWIA vary strongly
from one choice to the other. Since relativistic effects are so important
in this response, even in the plane--wave limit discussed here, it is likely
that experimental data may reveal such effects.

For the transverse responses $R^T$ and $R^{TT}$ the differences seen in
fig.~7 between CC1 and CC2 results are solely due to the negative--energy
projections (see also figs.~9,10,11). Again, these contributions 
($R^{T,TT}_C$, $R^{T,TT}_N$) are maximal for the CC1 choice. However, in
the case of $R^T$ their total contribution at the maximum is less than
$10\%$ for CC1 and less than $3\%$ for CC2.  Relatively much larger are
the contributions of the negative--energy projections to $R^{TT}$, although this
response is very small compared to $R^T$ and it is difficult to isolate
$R^{TT}$ from differential cross section measurements.

Focusing on the choices CC1$^{(0)}$ and CC2$^{(0)}$ that are closer to NCC2,
in all cases we show in figs.~10 and 11, respectively, the total $R^K$
responses and their $P$, $N$ and $C$ terms
to better appreciate the relative importance of the negative--energy
projections in each response function.  Clearly for the large responses
$R^L$ and $R^T$ the difference between the total relativistic responses
and their positive--energy projection is negligible. This in turn implies
that for these responses one cannot expect to see large effects in going
from PWIA to RPWIA, since also $N_{uu} \approx N_{uu}^{n.r.}$. Hence the
factorization limit of PWIA is practically preserved by RPWIA
in the $R^L$ and $R^T$ responses except when the CC1$^{(3)}$ prescription is
used for the current. On the contrary, the small response $R^{TT}$ and the
longitudinal transverse response $R^{TL}$ are very sensitive to
contributions from the negative--energy components, because the contributions
from positive--energy projections are relatively small. For these responses
the factorization limit of PWIA is badly broken even when the more
conservative CC2$^{(0)}$ choice is used (see fig.~11).

The breaking of this PWIA factorization limit in RPWIA  has important
consequences in the differential cross section. This is illustrated in
figs.~12--14 for different kinematical situations. Fig.~12 is for
kinematics I at two different electron scattering angles (left
$\theta_e=30^0$, right $\theta_e=150^0$), while figs.~13,14 are for
kinematics II and III at forward angle ($\theta_e=12.5^0$). 
We compare the fully relativistic results given by the prescriptions 
CC1$^{(0)}$ (thin--solid), CC2$^{(0)}$ (thick--solid),
CC1$^{(3)}$ (thin--dash) and CC2$^{(3)}$ (thick--dash)
to the non--relativistic
PWIA limit (dotted). The PWIA results in these figures have
been calculated from eq.~(\ref{eq1new}) with $N_b(p)$ as given by
$N_{uu}^{n.r.}(p)$ in eq.~(\ref{55new}) and with the CC1$^{(0)}$ choice.
It is important to stress here that these PWIA results practically coincide
with the positive--energy projection contribution of the total RPWIA
differential cross section in eq.~(\ref{eq1}). Indeed the difference
comes only from the difference between $N_{uu}(p)$ (eq.~(\ref{77p1})) and
$N_{uu}^{n.r.}(p)$ (eq.~(\ref{55new})), which is negligible, as seen in the
right--hand panel of fig.~1. Hence the differences between the dotted and
thin--solid (CC1$^{(0)}$) lines in figs.~12--14 are due to the contributions
from the negative--energy projections. Differences with thick--solid
(CC2$^{(0)}$), thick--dash (CC2$^{(3)}$), and thin--dash
(CC1$^{(3)}$) are also partly due to the effect of the negative--energy
contributions and partly due to the different current operator.

Clearly seen
in these figures is the fact that there are two important effects of the
negative--energy projections of the bound nucleon wave function:
A) A modification of the strength of the peak 
at low $p$ ($p < 300$ MeV), where data are
used to determine spectroscopic factors
(particularly important for kinematics I and for the prescriptions
CC1$^{(3)}$ and CC2$^{(3)}$);
B) A substantial modification of the
shape of the differential cross section, particularly in the high $p$
region. This latter effect is apparent in all of the different kinematical
situations considered. For instance in kinematics I one sees that even
in the case $\theta_e=150^0$, where the various prescriptions give
practically the same results in the low $p$ region and the RPWIA
practically coincides with PWIA, the high $p$ region ($300<p<500$ MeV)
shows important relativistic effects whose size depends on the choice of
the current operator.

We also point out that at high $p$ the cross section is
significatively higher for the CC1 choices.
The results shown in fig.~12 reflect the behaviour observed in fig.~7. A
detailed quantitative analyses of the role of the negative--energy projections
on the differential cross sections for different current choices at any
$p$--value and various kinematics observed in figs.~12--14 can also be
performed using the results shown in figs.~1,~5 and~6.

\section*{5. Summary and final remarks}

We have studied the relationship between relativistic and non--relativistic
treatments of the plane--wave impulse approximation to A(e,e$'$p)B
reactions by inserting the completeness relation into the relativistic
transition nuclear current. This allows one to separate the RPWIA differential
cross section into contributions from positive--energy projections and from
negative--energy projections of the bound nucleon wave function. This
separation, exhibited in eq.~(\ref{eq1}), clearly demonstrates that the
factorization limit of PWIA expressed by eq.~(\ref{int1}) (see also
eq.~(\ref{eq1new})) breaks down in RPWIA due to the presence of the
negative--energy projection that although small is non-zero for a
relativistic bound nucleon wave function. Typically, $N_{uv}$ is one order
of magnitude smaller than $N_{uu}$ and $N_{vv}$ is two orders of magnitude
smaller for $p < 300$ MeV. This is so for the $1p_{1/2}$ shell in
$^{16}O$ considered here, and we have observed a similar trend for various
shells in different nuclei~\cite{progress}. 

If we consider only the positive--energy projection of the relativistic
bound nucleon wave function we recover the PWIA factorized expression that
relates $A(e,e'p)B$ scattering to free electron-proton scattering through
the elementary differential cross-section $\sigma^{ep}=\sigma^{ep}_{uu}$.
However if we also take into account the negative--energy projection of the
relativistic bound nucleon wave function other single--nucleon components
$\sigma^{ep}_{uv}$ and $\sigma^{ep}_{vv}$ appear that are not present
in the case of electron scattering from a free nucleon (or antinucleon).
It is in this sense that we speak of lack of factorization in RPWIA.
Obviously, the extent to which factorization breaks down is measured by
the relative importance of the contributions from the negative--energy
projections. 

We have studied the relative importance of these contributions with
different choices of the current operators and of the kinematics. A
quantitative study has been presented of ---both single-nucleon 
and total--- response functions and 
differential cross sections for different choices
of the current operator and of the kinematics, focussing on the case of a
$1p_{1/2}$ bound nucleon in $^{16}O$. The main outcome of this study can be
summarized as follows.

The role of the negative--energy component in the total differential cross
section is to cause a reduction at low $p$ ($p<300$ MeV) and an increase at
high $p$ ($p\gsim 300$ MeV). This affects both the value of the spectroscopic
factors and the shape of the extracted momentum distribution
$\tilde{\rho}(p)$~\cite{ud93,ud95,ud96}. 

The reduction at low $p$ is generally
small (less than a 15\% for the NCC1, NCC2, CC1$^{(0)}$ and CC2$^{(0)}$
choices)
and depends mainly on the kinematics. The largest reduction
is found at small $q$
values ($q< 500$ MeV) and at forward angles for the current choice CC1$^{(3)}$.
The sizeable difference at low $p$ and $q$ (forward angles) observed
between the predictions of the CC1$^{(0)}$ and CC1$^{(3)}$ 
choices can be traced back to the difference
between the longitudinal currents in eqs.~(\ref{c18a},\ref{c18b}), and it
is also present in the positive--energy projections. Actually a sizeable
difference at these kinematics is also observed between
CC1$^{(0)}$ and CC1$^{(3)}$
results (as well as between CC2$^{(0)}$ and CC2$^{(3)}$ results) in 
the single--nucleon component $\sigma_{uu}$.

At low $p$, the positive--energy projections are dominant and the RPWIA
results do not differ so much from the PWIA ones; however at $p> 300$ MeV
the negative--energy projections play a more important role and the
differences between current choices are enhanced. Indeed the 
single--nucleon components $\sigma_{uv}$ and $\sigma_{vv}$ in general depend
much more on the choice of the current operator (CC1 or CC2) than does the
positive--energy projection
$\sigma_{uu}$ and this is mainly seen in the total differential cross
sections  at high $p$.

As a general rule, the NCC2 and CC2$^{(0)}$ choices minimize the effect of
the negative--energy 
components, while all choices corresponding to the CC1 operator tend to
enhance the contributions of the negative--energy components.
The analyses of the nuclear current components in coordinate space show
that, compared to CC2, the CC1 choices give more weight to the dynamical
enhancement of the lower components that appear multiplied by the {\em
large} $S+V$ potential.

The response functions $R^{TL}$ and $R^{TT}$ are particularly sensitive to
the combined effect of the negative--energy component and the choice of the
current operator, because the cross term $R^{TL}_C$ ($R^{TT}_C$) can be
larger than the positive--energy projection $R^{TL}_P$ ($R^{TT}_P$). As a 
net result one can find up to a factor of 3 difference between predictions
of various current choices. While there is basically no difference between
NCC2 and CC2$^{(0)}$ predictions, the choices NCC1, CC1$^{(0)}$ and
CC2$^{(3)}$  predict an
$R^{TL}$ response that is about 
$40\%$ larger, and the CC1$^{(3)}$ choice predicts a
three times larger $R^{TL}$ response. These large effects can be traced
back to the difference between the CC1 and CC2 current 
operators in eq.~(\ref{c21b}). 
Clearly the $(\xslash{P}-M)$ term that has no effect in the
positive--energy projection $R^{TL}_P$ has an important effect in the
crossed terms $R^{TL}_C$ and in the 
negative--energy term $R^{TL}_N$. Obviously the $(\xslash{P}-M)$
term also causes differences between CC1 and CC2 predictions 
in the $R_C$ and $R_N$ terms of other
response functions, but it is in the case of the TL and TT responses where the
net effect is maximum because of the small values of the positive--energy
projections.

The effect in $R^{TL}$ is particularly interesting because
this response can be more easily measured than $R^{TT}$ and data on
$R^{TL}$ are already available.
Whereas for most of the current prescriptions (NCC1, NCC2, CC1$^{(0)}$ and
CC2$^{(0)}$) and at low $p$ values ($p\lsim 400$ MeV) the relativistic
and non--relativistic results for the longitudinal 
and transverse responses differ at most by a 15\%, in the case of the
longitudinal-transverse ($R^{TL}$) response the relativistic prediction is
at least 
a 50\% larger than the non--relativistic one.
It is important to point out that there is experimental evidence of the
fact that a different 
spectroscopic factor is needed for the $R^{TL}$ response than for
the $R^{L}$ and $R^{T}$ ones. This experimental evidence follows from a 
non--relativistic analyses of the data~\cite{spal93,bul93}. Although
considering other effects such
as two--body meson--exchange currents 
can explain part of this difference~\cite{slu94},
it is clear that relativity plays a very important role in understanding
this response, and it will be very interesting
to see to what extent such discrepancies may be related to the
relativistic effects found here~\cite{progress}.

\subsection*{Acknowledgements}
This work is supported in part by a NATO Collaborative Research Grant Number
940183, in part by DGICYT (Spain) under Contract Nos. PB/95--0123 and
PB/95--0533--A and in part by funds provided by the US Department of Energy
(D.O.E.) under cooperative agreement \#DE--FC01--94ER40818.

\section*{Appendix A}
\begin{itemize}

\item {\large \bf Free Dirac Spinors}

\begin{enumerate}

We follow the conventions of Bjorken and Drell~\cite{bd}. In particular,
the free Dirac spinors are

\beqa
u({\bf p},s)&=&\sqrt{\frac{\overline{E}+M}{2M}}
		\left(\begin{array}{@{\hspace{0pt}}c@{\hspace{0pt}}}
		\chi^s \\
		\frac{{\bd \sigma}\cdot{\bd p}}{\overline{E}+M}
		\chi^s\end{array}\right)
\\
v({\bf p},s)&=&\sqrt{\frac{\overline{E}+M}{2M}}
		\left(\begin{array}{@{\hspace{0pt}}c@{\hspace{0pt}}}
		\frac{{\bd \sigma}\cdot{\bd p}}{\overline{E}+M}
		\chi^s \\
		\chi^s\end{array}\right)
\eeqa
where $\overline{E}$ denotes $\overline{E}=\sqrt{M^2+{\bd p}^2}$.

\item Normalization
\beq
\overline{u}({\bd p},s)u({\bd p},s)=-\overline{v}({\bd p},s)v({\bd p},s)=1
\,\,\, .
\eeq

\item Completeness relation

\beq
\sum_s\left[ u_\alpha ({\bd p},s)\overline{u}_\beta({\bd p},s)-
		v_\alpha ({\bd p},s)\overline{v}_\beta({\bd p},s)\right]
	=\delta_{\alpha \beta} \,\,\, .
\eeq
\end{enumerate}

\item {\large \bf Relativistic Bound Nucleon Wave Function}

The relativistic wave function for the bound nucleon in momentum space
is given by
\beq
\Psi_\kappa^m({\bd p})=
	\frac{1}{(2\pi)^{3/2}}\int d{\bd r}e^{-i{\bd p}\cdot{\bd r}}
	\Psi_\kappa^m({\bd r})
	=(-i)^\ell
	\left(\begin{array}{@{\hspace{0pt}}c@{\hspace{0pt}}}
		g_\kappa(p) \\
		 S_\kappa f_\kappa (p)\frac{{\bd \sigma}\cdot{\bd p}}{p}
		\end{array}\right)\Phi_\kappa^m(\Omega_p)
\eeq
with $S_\kappa=\kappa/|\kappa|$, $j=|\kappa|-\frac{1}{2}$ and the quantum
number $\ell$ given by the relation
\beq
\ell=\left\{\begin{array}{@{\hspace{0pt}}c@{\hspace{0pt}}}
		\kappa \,\,\,\, for \,\,\,\, \kappa>0 \\
		 -\kappa-1 \,\,\,\, for \,\,\,\, \kappa<0
		\end{array}\right\}
\eeq
The function $\Phi_\kappa^m(\Omega_p)$ is given by
\beq
\Phi_\kappa^m(\Omega_p)=\langle \hat{p}|\ell \frac{1}{2} j m \rangle =
	\sum_{\mu s}\langle \ell\mu\frac{1}{2} s |jm\rangle
	Y_\ell^\mu(\Omega_p)\chi^s =
	\sum_s \chi^s \langle s|\Phi_\kappa^m \rangle
\eeq
and satisfies the relation 
\beq
\Phi_{-\kappa}^m (\Omega_p)=-\frac{{\bd \sigma}\cdot {\bd p}}
		{p}\Phi_\kappa^m(\Omega_p) \,\,\, .
\label{apa1}
\eeq

The radial functions $g_\kappa$ and $f_\kappa$ in momentum space are
obtained from the respective functions in coordinate space
\beqa
g_\kappa (p) &=& \sqrt{\frac{2}{\pi}}
	\int_0^\infty r^2 dr g_\kappa(r) j_\ell (pr) 
		\\
f_\kappa (p) &=& \sqrt{\frac{2}{\pi}}
	\int_0^\infty r^2 dr f_\kappa(r) j_{\overline{\ell}} (pr) 
\eeqa
with $j_\ell(pr)$ the Ricati--Bessel functions and
$\ell =\kappa$, $\overline{\ell}=\kappa -1$ for $\kappa > 0$;
$\ell =|\kappa | -1$, $\overline{\ell}=|\kappa |$ for $\kappa < 0$.
The radial functions in coordinate space $g_\kappa (r)$ and $f_\kappa (r)$
satisfy the Dirac equation
\beqa
\frac{df_\kappa}{dr} &=&\frac{\kappa -1}{r}f_\kappa -\left[
	E-M-U_S-U_V\right]g_\kappa
\\
\frac{dg_\kappa}{dr} &=&-\frac{\kappa -1}{r}g_\kappa -\left[
	E-M+U_S-U_V\right]f_\kappa
\eeqa
with $U_S$ ($U_V$) the scalar (vector) potentials that describe the
target nucleus [12,13], and the normalization is
\beq
\int r^2 dr \left(g_\kappa ^2(r)+f_\kappa ^2(r)\right) =
\int p^2 dp \left(g_\kappa ^2(p)+f_\kappa ^2(p)\right) = 1 \,\,\, .
\eeq

Hence the relativistic (vector) density in momentum space normalized to 1
is 
\beq
N_b^r(p) = \frac{1}{\hat{j}^2}\sum_m \Psi_\kappa^{m +}\Psi_\kappa^m =
	\frac{\left(g_\kappa ^2(p)+f_\kappa ^2(p)\right)}{4\pi} \,\,\, .
\eeq

\end{itemize}

\section*{Appendix B}

In this appendix we present in detail the algebra needed in order to evaluate
the positive and negative energy projection contributions to the hadronic tensor
(see eq.~(\ref{24})).
Let us start with the contribution from the
positive--energy projections. 
It is given by (for simplicity we suppress the index $b$ on the bound
proton wave function and quantum numbers)

\beqa
W^{\mu\nu}_P 
&\equiv &\frac{2}{2j+1}\sum_{m}\sum_{s_N}\langle\hat{J}^\mu \rangle^*_u
	\langle\hat{J}^\nu \rangle_u
\nonumber \\
&=& \frac{2}{2j+1}\sum_{m}\sum_{ss'}
	\left[\overline{u}({\bd p},s)\,\,\Psi_\kappa^m({\bd p})\right]^*
	\left[\overline{u}({\bd p},s')\,\,\Psi_\kappa^m({\bd p})\right]
\nonumber \\
&\times&
	\sum_{s_N}\left[\overline{u}({\bd p}_N,s_N)\hat{J}^\mu
			u({\bd p},s)\right]^*
		\left[\overline{u}({\bd p}_N,s_N)\hat{J}^\nu
			u({\bd p},s')\right] \,\,\, .
\label{25}
\eeqa
Introducing the tensor ${\cal W}^{\mu\nu}_{s's}$ defined as
\beq
{\cal W}^{\mu\nu}_{s's}=
\sum_{s_N}\left[\overline{u}({\bd p}_N,s_N)\hat{J}^\mu
			u({\bd p},s)\right]^*
		\left[\overline{u}({\bd p}_N,s_N)\hat{J}^\nu
			u({\bd p},s')\right]
\label{26}
\eeq
and using the result obtained for the coefficient
$\left[\overline{u}({\bd p},s) \,\, \Psi_\kappa^m({\bd p})\right]$ as given
in eq.~(\ref{18}), we can write
\beq
W^{\mu\nu}_P=
	\frac{2}{2j+1}\left(\frac{\overline{E}+M}{2M}\right)
		|\alpha_\kappa(p)|^2
	 \sum_{ss'}{\cal W}^{\mu\nu}_{s's}
	\sum_{m}
	\langle s|\Phi_\kappa^m\rangle^\ast
	\langle s'|\Phi_\kappa^m\rangle \,\,\, .
\label{27}
\eeq
Now we can make use of the general relation given by eq.~(\ref{62pp}) 
to get the result shown in eq.~(\ref{29pc}).

In the case of the contribution from negative--energy projections, we have
\beqa
W^{\mu\nu}_N
&\equiv &\frac{2}{2j+1}\sum_m\sum_{s_N}\langle\hat{J}^\mu \rangle^*_v
	\langle\hat{J}^\nu \rangle_v
\nonumber \\
&=& \frac{2}{2j+1}\sum_m\sum_{ss'}
	\left[\overline{v}({\bd p},s)\,\,\Psi_\kappa^m({\bd p})\right]^*
	\left[\overline{v}({\bd p},s')\,\,\Psi_\kappa^m({\bd p})\right]
\nonumber \\
&\times &
	\sum_{s_N}\left[\overline{u}({\bd p}_N,s_N)\hat{J}^\mu
			v({\bd p},s)\right]^*
		\left[\overline{u}({\bd p}_N,s_N)\hat{J}^\nu
			v({\bd p},s')\right] \,\,\, .
\label{31}
\eeqa

Introducing the tensor ${\cal Z}^{\mu\nu}_{s's}$ given by
\beq
{\cal Z}^{\mu\nu}_{s's}=
\sum_{s_N}\left[\overline{u}({\bd p}_N,s_N)\hat{J}^\mu
			v({\bd p},s)\right]^*
		\left[\overline{u}({\bd p}_N,s_N)\hat{J}^\nu
			v({\bd p},s')\right]
\label{32}
\eeq
and using the results given by eqs.~(\ref{21},\ref{apa1}),
we can write
\beqa
W^{\mu\nu}_N &=&
	\frac{2}{2j+1}\left(\frac{\overline{E}+M}{2M}\right)
		|\beta_\kappa(p)|^2 	\sum_{ss'}{\cal Z}^{\mu\nu}_{s's}
	\sum_m \langle s|\Phi_{-\kappa}^m \rangle^\ast
	\langle s' |\Phi_{-\kappa}^m \rangle
\nonumber \\
	&= & \frac{2}{2j+1}\left(\frac{\overline{E}+M}{2M}\right)
		|\beta_\kappa(p)|^2\frac{1}{p^2}
	\sum_{ss'}{\cal Z}^{\mu\nu}_{s's}
\nonumber \\
&\times& \sum_{\delta \delta'}
	\left[\chi^+_s({\bd \sigma}\cdot{\bd p})\chi^\delta\right]^\ast 
	\left[\chi^+_{s'}({\bd \sigma}\cdot{\bd p})\chi^{\delta'}\right]
	\sum_m 
	\langle \delta |\Phi_\kappa^m \rangle^\ast
	\langle \delta' |\Phi_\kappa^m \rangle \,\,\, .
\nonumber \\
& & 
\label{33}
\eeqa

Using the relation given by eq.~(\ref{62pp}), we get
\beq 
W^{\mu\nu}_N =
	\left(\frac{1}{8\pi}\right)\left(\frac{\overline{E}+M}{M}\right)
	\left|\beta_\kappa(p)\right|^2\frac{1}{p^2}
	\sum_{ss'}{\cal Z}^{\mu\nu}_{s's}\sum_{\delta}
	\left[\chi^+_s({\bd \sigma}\cdot{\bd p})\chi^\delta\right]^\ast 
	\left[\chi^+_{s'}({\bd \sigma}\cdot{\bd p})\chi^{\delta}\right]
	\,\,\, ,
\label{34}
\eeq
where the sum over the index $\delta$ is simply
\beq
\sum_\delta
\left[\chi^+_s({\bd \sigma}\cdot{\bd p})\chi^\delta\right]^\ast 
	\left[\chi^+_{s'}({\bd \sigma}\cdot
	{\bd p})\chi^{\delta}\right]= 	p^2\delta_{ss'}
\label{35}
\eeq
and the final result reduces to the expression given in eq.~(\ref{36pc}).
Note that the single--nucleon tensor
${\cal Z}^{\mu\nu}\equiv \sum_s{\cal Z}^{\mu\nu}_{ss}$ can be easily
calculated using trace techniques (eq.~(\ref{37p}).)

Let us proceed now to the evaluation of the cross term containing products
of both positive and negative energy projections. Proceeding as in the
previous cases and using eq.~(\ref{63pp}) we can write
\beqa 
W^{\mu\nu}_C &=&
\frac{-2}{2j+1}\sum_m\sum_{s_N}\left (\langle \hat{J}^\mu \rangle_u^\ast
	\langle \hat{J}^\nu \rangle _v +
	\langle \hat{J}^\mu \rangle_v^\ast
	\langle \hat{J}^\nu \rangle _u \right)
\nonumber \\
&=&
\left(\frac{1}{8\pi}\right)\left(\frac{\overline{E}+M}{M}\right)
\alpha_\kappa(p)
	\beta_\kappa(p)\frac{1}{p}\sum_{ss'}
	\left({\cal I}^{\mu\nu}_{s's}+ {\cal I}^{\nu\mu\ast}_{ss'} \right)
	\left[\chi^+_{s'}({\bd \sigma}\cdot {\bd p})\chi^s
		\right] \,\,\, ,
\label{38}
\eeqa
where we have 
introduced a new single--nucleon tensor defined as follows,
\beq
{\cal I}^{\mu\nu}_{s's}=\sum_{s_N}
\left[\overline{u}({\bd p}_N,s_N)\hat{J}^\mu u({\bd p},s) \right]^\ast
\left[\overline{u}({\bd p}_N,s_N)\hat{J}^\nu v({\bd p},s') \right] \,\,\, .
\label{40}
\eeq

From general properties of the $\gamma$ matrices and making use of the
trace techniques \cite{bd}, we can express the tensor 
${\cal I}^{\mu\nu}_{s's}$ as
\beq
{\cal I}^{\mu\nu}_{s's}=\frac{1}{8M^2}\Tr\left\{\gamma_5(\sla{P}+M)
	(\delta_{ss'}+\gamma_5\sla{\varphi}_{s's})\overline{J}^\mu
	(\sla{P}_N+M)J^\nu\right\} \,\,\, ,
\label{49} 
\eeq
where the pseudovector $\varphi^\mu_{ss'}$ (see
ref.~\cite{dp} for details) reduces to the four--spin of the bound nucleon
$S^\mu_L$ in the diagonal case, $s'=s$. 

It is clear from the above
expression that the off--diagonal contributions are purely symmetric
and therefore one may write
\beq
{\cal I}^{\mu\nu}_{s's}={\cal S}^{\mu\nu}_{s's}+
		i\delta_{ss'}{\cal A}^{\mu\nu} \,\,\, ,
\label{53}
\eeq
where ${\cal A}^{\mu\nu}$ is antisymmetric
under $\mu\leftrightarrow \nu$ and real, whereas ${\cal S}^{\mu\nu}_{s's}$
is symmetric under
$\mu\leftrightarrow \nu$, real for diagonal terms and in general
complex for off--diagonal terms. These properties combined with the relation
\beq
\left[\chi^+_{s}({\bd \sigma}\cdot{\bd p})\chi^s\right]=
-\left[\chi^+_{-s}({\bd \sigma}\cdot{\bd p})\chi^{-s}\right]
\label{58}
\eeq
allow one to write finally,
\beq
W^{\mu\nu}_C=
 \left(\frac{-1}{8\pi}\right)\left(\frac{\overline{E}+M}{M}\right)\alpha_\kappa(p)
	\beta_\kappa(p)\sum_{ss'}{\cal R}^{\mu\nu}_{s's}
	\left[\chi^+_{s'}(\frac{{\bd \sigma}\cdot {\bd p}}{p})\chi^s
		\right]
\label{63}
\eeq
with
\beq
{\cal R}^{\mu\nu}_{s's}=\frac{1}{4M}\Tr \left[
	\sla{\varphi}_{s's}\overline{J}^\mu(\sla{P}_N+M)J^\nu
		\right] \,\,\, .
\label{64}
\eeq
It can be proved that $\sum_{ss'}{\cal R}^{\mu\nu}_{s's}
	\left[\chi^+_{s'}(\frac{{\bd \sigma}\cdot {\bd p}}{p})\chi^s
		\right]$ is just a trace and can be simply written as
\beq
\sum_{ss'}{\cal R}^{\mu\nu}_{s's}
	\left[\chi^+_{s'}(\frac{{\bd \sigma}\cdot {\bd p}}{p})\chi^s
		\right]=
\Tr \left[\overline{J}^\mu\left(\frac{\sla{P}_N+M}{2M}\right)
	J^\nu \gamma^0 \frac{{\bd \gamma} \cdot {\bd p}}{p}
	\frac{\sla{P}}{M}\right] = 2{\cal N}^{\mu\nu}
\eeq
with the tensor ${\cal N}^{\mu\nu}$ as introduced in section 2.3
(see eq.~(\ref{37pp})).

\section*{Appendix C}

In this appendix we 
give the explicit expressions of the single--nucleon tensors
${\cal W}^{\mu\nu}$, ${\cal Z}^{\mu\nu}$ and ${\cal R}^{\mu\nu}$ for the
two current operators CC1 and CC2 in eqs.~(\ref{c5},\ref{c4}).

The following expressions for the various single--nucleon
tensors are obtained:
\begin{itemize}

\item with $\hat{J}^\mu_{CC1}$ current operator

\beqa
& & M^2{\cal W}^{\mu\nu}=\left(F_1+F_2\right)^2
\left(\overline{P}^\mu P_N^\nu +
\overline{P}^\nu P_N^\mu+\frac{\overline{Q}^2}{2}g^{\mu\nu}\right)
\nonumber \\
&-&\left[F_2(F_1+F_2)-F_2^2\left(\frac{1}{2}-
	\frac{\overline{Q}^2}{8M^2}\right)\right]
	(\overline{P}+P_N)^\mu (\overline{P}+P_N)^\nu
\label{d3}
\eeqa

\beqa
& & M^2{\cal Z}^{\mu\nu}=\left(F_1+F_2\right)^2
\left(\overline{P}^\mu P_N^\nu +
\overline{P}^\nu P_N^\mu-\frac{(\overline{P}+P_N)^2}{2}g^{\mu\nu}\right)
\nonumber \\
&-&\frac{F_2^2}{8M^2}\overline{Q}^2(\overline{P}+P_N)^\mu 
(\overline{P}+P_N)^\nu
	+F_2(F_1+F_2)\left(P_N^\mu P_N^\nu-
		\overline{P}^\nu \overline{P}^\mu \right)
\label{d4}
\eeqa

\beqa
M{\cal R}^{\mu\nu}&=&(F_1+F_2)^2
\left(S_L^\mu P_N^\nu +S_L^\nu P_N^\mu-P_N\cdot S_L g^{\mu\nu}\right)
\nonumber \\
&+&\frac{F_2^2}{4M^2}P_N\cdot S_L 
	(\overline{P}+P_N)^\mu (\overline{P}+P_N)^\nu
\nonumber \\
&-&\frac{F_2}{2}(F_1+F_2)\left(\frac{}{}
S_L^\mu (\overline{P}+P_N)^\nu+
		S_L^\nu (\overline{P}+P_N)^\mu\right)
\label{d5}
\eeqa

\item with $\hat{J}^\mu_{CC2}$ current operator

\beqa
M^2 {\cal W}^{\mu\nu} &=& F_1^2 \left(
	\overline{P}^\mu P_N^\nu + \overline{P}^\nu P_N^\mu+
	\frac{\overline{Q}^2}{2}g^{\mu\nu}\right) +
F_1F_2\left(Q\cdot\overline{Q}\,\,g^{\mu\nu}-
	\frac{\overline{Q}^\mu Q^\nu+\overline{Q}^\nu Q^\mu}{2}\right) 
\nonumber \\
&+&\frac{F_2^2}{4M^2}\left[ \frac{}{}
P_N\cdot Q \left(\overline{P}^\mu Q^\nu 
		+\overline{P}^\nu Q^\mu\right) +
\overline{P}\cdot Q \left(P_N^\mu Q^\nu +P_N^\nu Q^\mu\right) \right. 
\nonumber \\
&-&\left. Q^2\left(P_N^\mu \overline{P}^\nu+P_N^\nu \overline{P}^\mu\right)-
\left(2M^2-\frac{\overline{Q}^2}{2}\right)Q^\mu Q^\nu \right. 
\nonumber \\
&+&\left. g^{\mu\nu}\left(2M^2Q^2-\frac{Q^2\overline{Q}^2}{2}-
	2P_N\cdot Q \overline{P}\cdot Q\right) \right]
\label{d6}
\eeqa

\beqa
& & M^2{\cal Z}^{\mu\nu}  = F_1^2 \left(
\overline{P}^\mu P_N^\nu + \overline{P}^\nu P_N^\mu-
	\frac{(\overline{P}+P_N)^2}{2}g^{\mu\nu}\right)
\nonumber \\
& +&
F_1F_2\left(
	\frac{Q^\mu (\overline{P}+P_N)^\nu+Q^\nu (\overline{P}+P_N)^\mu}{2}-
Q\cdot (\overline{P}+P_N) g^{\mu\nu} \right)
\nonumber \\
&+&\frac{F_2^2}{4M^2}\left[ \frac{}{}
	P_N\cdot Q \left(\overline{P}^\mu Q^\nu +\overline{P}^\nu
Q^\mu\right) +
\overline{P}\cdot Q \left(P_N^\mu Q^\nu +P_N^\nu Q^\mu\right) \right. 
\nonumber \\
&-&\left. Q^2\left(P_N^\mu \overline{P}^\nu+P_N^\nu \overline{P}^\mu\right)+
\frac{\overline{Q}^2}{2}Q^\mu Q^\nu 
- g^{\mu\nu}\left(\frac{Q^2\overline{Q}^2}{2}+
	2P_N\cdot Q \overline{P}\cdot Q\right) \right]
\label{d7}
\eeqa

\beqa
& & M{\cal R}^{\mu\nu} = F_1^2 \left(S_L^\mu P_N^\nu + S_L^\nu P_N^\mu-
	P_N\cdot S_L g^{\mu\nu}\right)
\nonumber \\
& +&
\frac{F_1F_2}{2}\left(
	S_L^\mu Q^\nu+S_L^\nu Q^\mu-
2Q\cdot S_L g^{\mu\nu} \right)
\nonumber \\
&+&\frac{F_2^2}{4M^2}\left[ \frac{}{}
P_N\cdot Q \left(S_L^\mu Q^\nu +S_L^\nu
Q^\mu\right) +
S_L\cdot Q \left(P_N^\mu Q^\nu +P_N^\nu Q^\mu\right) \right. 
\nonumber \\
&-&\left. Q^2\left(P_N^\mu S_L^\nu+P_N^\nu S_L^\mu\right)-
P_N\cdot S_L Q^\mu Q^\nu
+ g^{\mu\nu}\left(Q^2 P_N\cdot S_L-
	2P_N\cdot Q S_L\cdot Q\right) \right]
\nonumber \\
\label{d8}
\eeqa
with $A\cdot B\equiv A_\mu B^\mu$. 

\end{itemize}

Finally, we show in table I
the components of the spin four--vector $S_L^\mu$ in the laboratory frame.

\begin{table}[h]
\vspace{1cm}
\begin{center}
\begin{tabular}{||c||c|c|c|c||} \hline

($\theta_R,\phi_R$)   & $\mu=0$ & $\mu=1$ & $\mu=2$ & $\mu=3$
					\\ \hline
    &       &      &  & \\
 $(0,0)$    &   $\chi'$ & $\frac{\chi\chi'}{\overline{\gamma}+1}$ 
& 0  &  $1+\frac{\chi'^2}{\overline{\gamma}+1}$  \\ 
    &   &    &      &   \\
$(\frac{\pi}{2},0)$ & $\chi\cos\phi$  &  
$\left(1+\frac{\chi^2}{\overline{\gamma}+1}\right)\cos\phi$  & 
$-\sin\phi$ & $\frac{\chi\chi'}{\overline{\gamma}+1}\cos\phi$  \\
 &    &       &      &   \\
$(\frac{\pi}{2},\frac{\pi}{2})$ & $\chi\sin\phi$  & 
$\left(1+\frac{\chi^2}{\overline{\gamma}+1}\right)\sin\phi$  & 
$\cos\phi$ & $\frac{\chi\chi'}{\overline{\gamma}+1}\sin\phi$ \\
		\hline 
\end{tabular}
\end{center}
\caption{Components of the spin four vector
$S^\mu_L(\theta_R,\phi_R)$. The notation
$\overline{\gamma}\equiv \frac{\overline{E}}{M}$;
$\chi\equiv \frac{p_N}{M}\sin\theta_N=\frac{p}{M}\sin\theta$;
$\chi'\equiv \frac{p}{M}\cos\theta=\frac{p_N}{M}\cos\theta_N-2\kappa$ and
$\kappa \equiv \frac{q}{2M}$ has been introduced}
\end{table}

\newpage
\noindent
\section*{Figure captions}
\begin{enumerate}
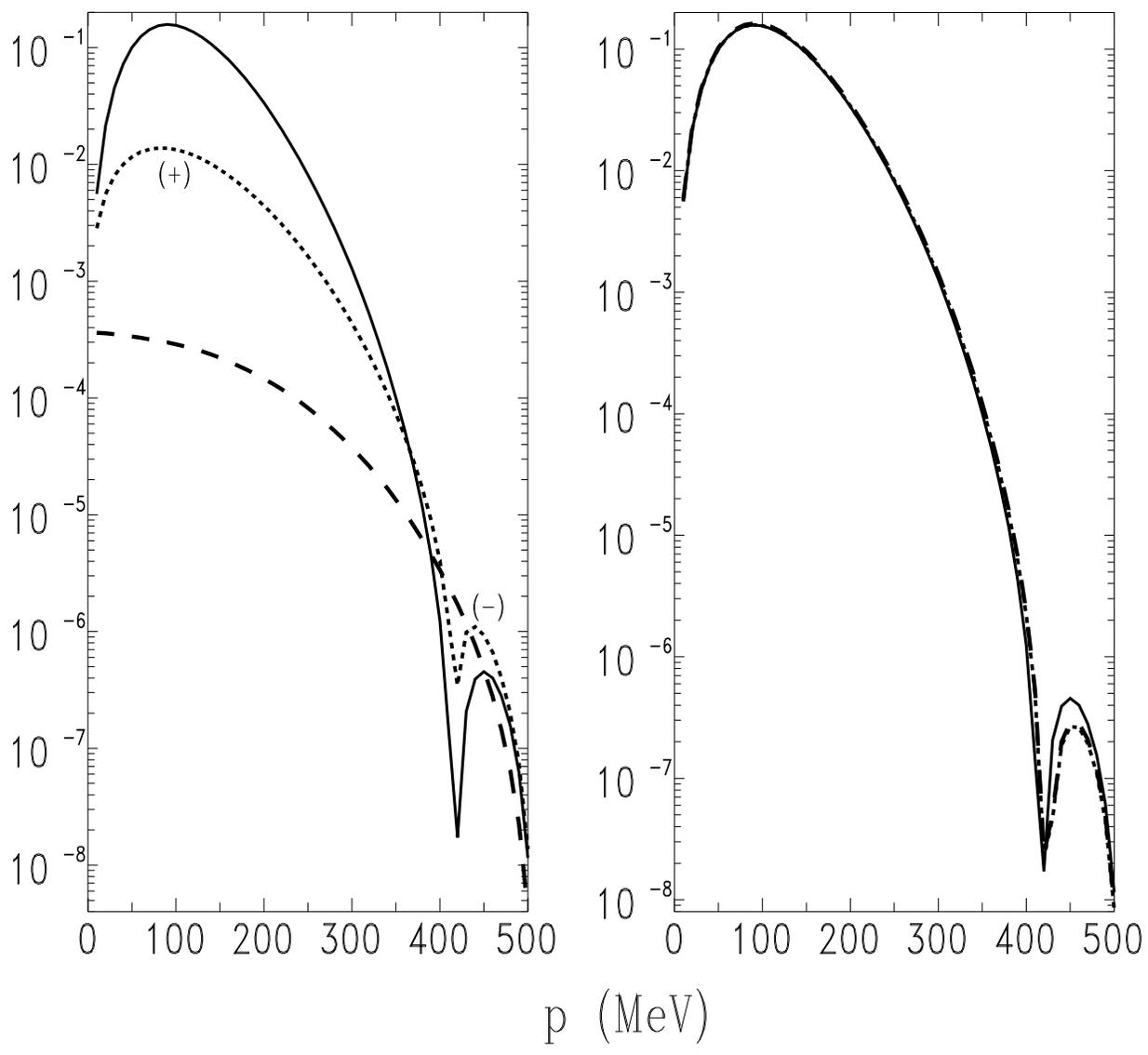
\item[Figure 1:]
Left panel: projection components of the momentum
distribution (in units of fm$^3$):
$N_{uu}(p)$ (solid), $N_{uv}(p)$ (dotted) and
$N_{vv}(p)$ (dashed). Right panel: $N_{uu}(p)$ (solid), $N_{uu}^{(0)}(p)$
(dotted) and $N_{uu}^{n.r.}(p)$ (dashed) (see text for details).

\item[Figure 2a:]
Projection components of the single--nucleon response functions,
${\cal R}^K_{uu}$, ${\cal R}^K_{uv}$ and ${\cal R}^K_{vv}$.
Kinematics I have been chosen and results for the
pure longitudinal (top) and interference longitudinal--tranverse (bottom)
responses are shown. Thick lines correspond to prescriptions that use the
CC2 current operator and thin lines correspond to the CC1 operator.
Prescriptions shown are: NCC2 (NCC1) (solid lines), CC2$^{(0)}$
(CC1$^{(0)}$) (short--dash lines) and CC2$^{(3)}$ (CC1$^{(3)}$)
(long--dash lines).

\item[Figure 2b:]
Same as fig.~2a, except that now for the pure transverse (top) and
tranverse--transverse interference (bottom) single--nucleon responses.
Note that here all CC2 (CC1) prescriptions collapse into a single 
thick (thin) solid curve.

\item[Figure 3a:]
Same as fig.~2a, except that now for kinematics II (see text).

\item[Figure 3b:]
Same as fig.~2b, except that now for kinematics II (see text).

\item[Figure 4a:]
Same as fig.~2a, except that now for kinematics III (see text).

\item[Figure 4b:]
Same as fig.~2b, except that now for kinematics III (see text).

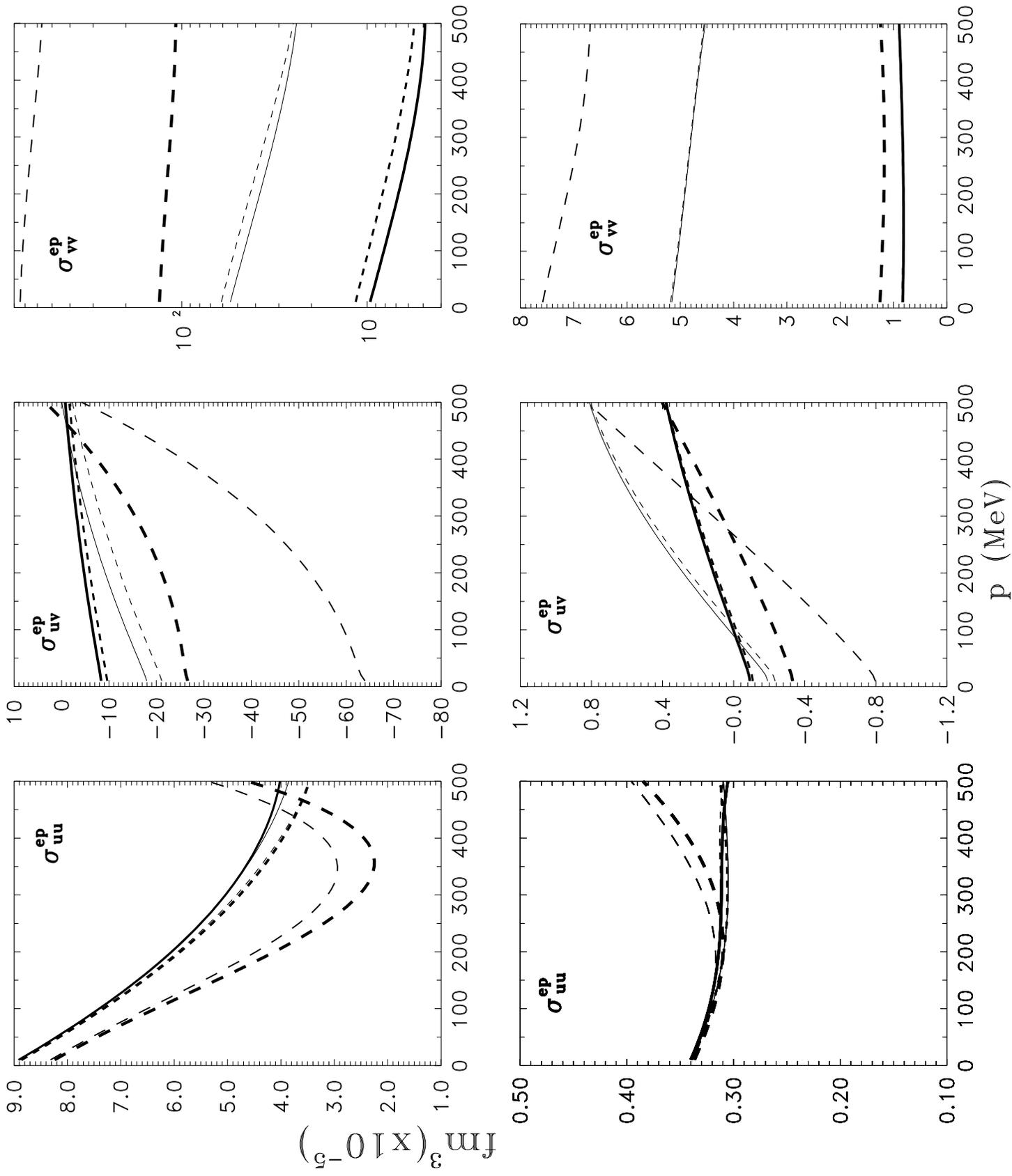
\item[Figure 5:]
Projections of the single--proton cross section as given by 
eqs.~(\ref{eq2a}-\ref{eq2c}).
Results correspond to kinematics I and $\phi_N=0^0$ (in-plane).
Forward-angle electron scattering ($\theta_e=30^0$) 
(top panel) and backward ($\theta_e=150^0$) (bottom) have been
chosen here. The labels are as in fig.~2a.

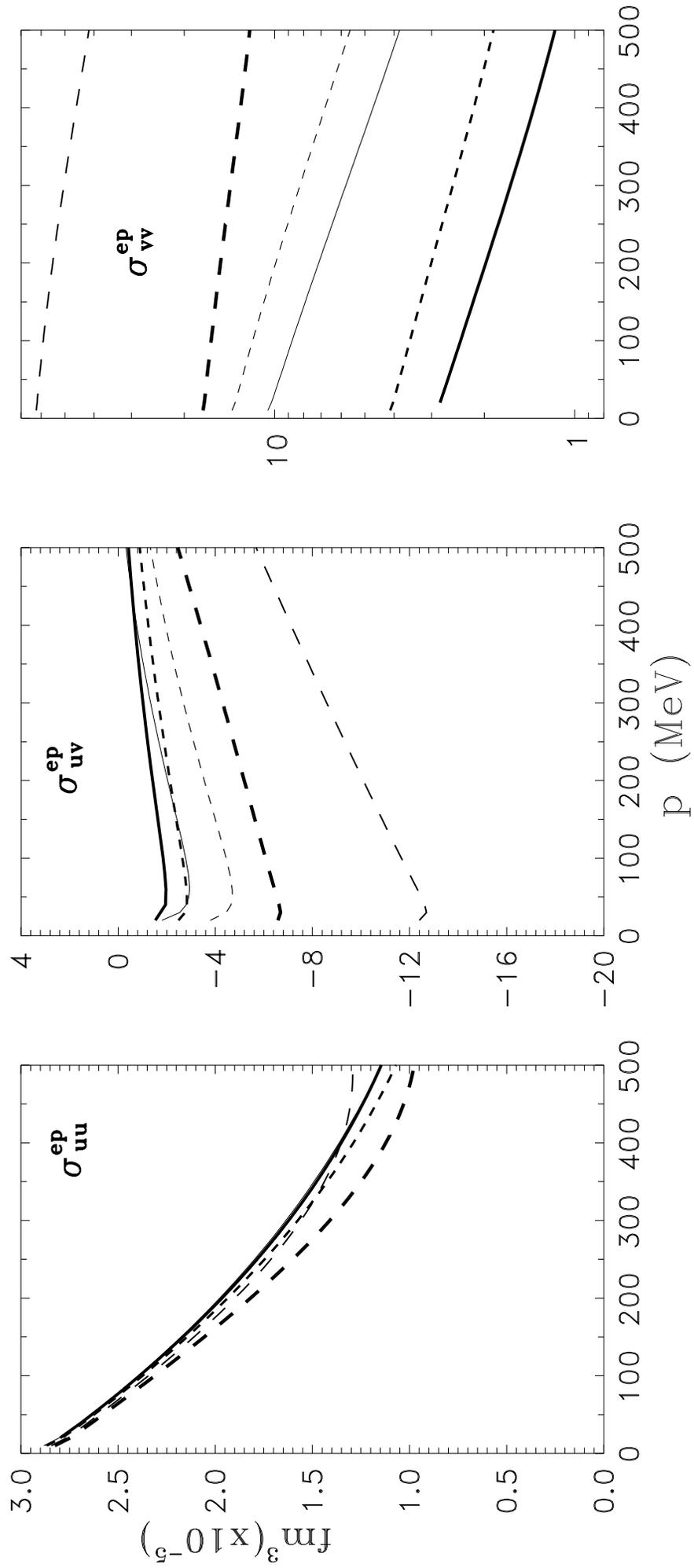
\item[Figure 6:]
Same as fig.~5, except now for kinematics II and
forward-angle electron scattering ($\theta_e=12.5^0$).

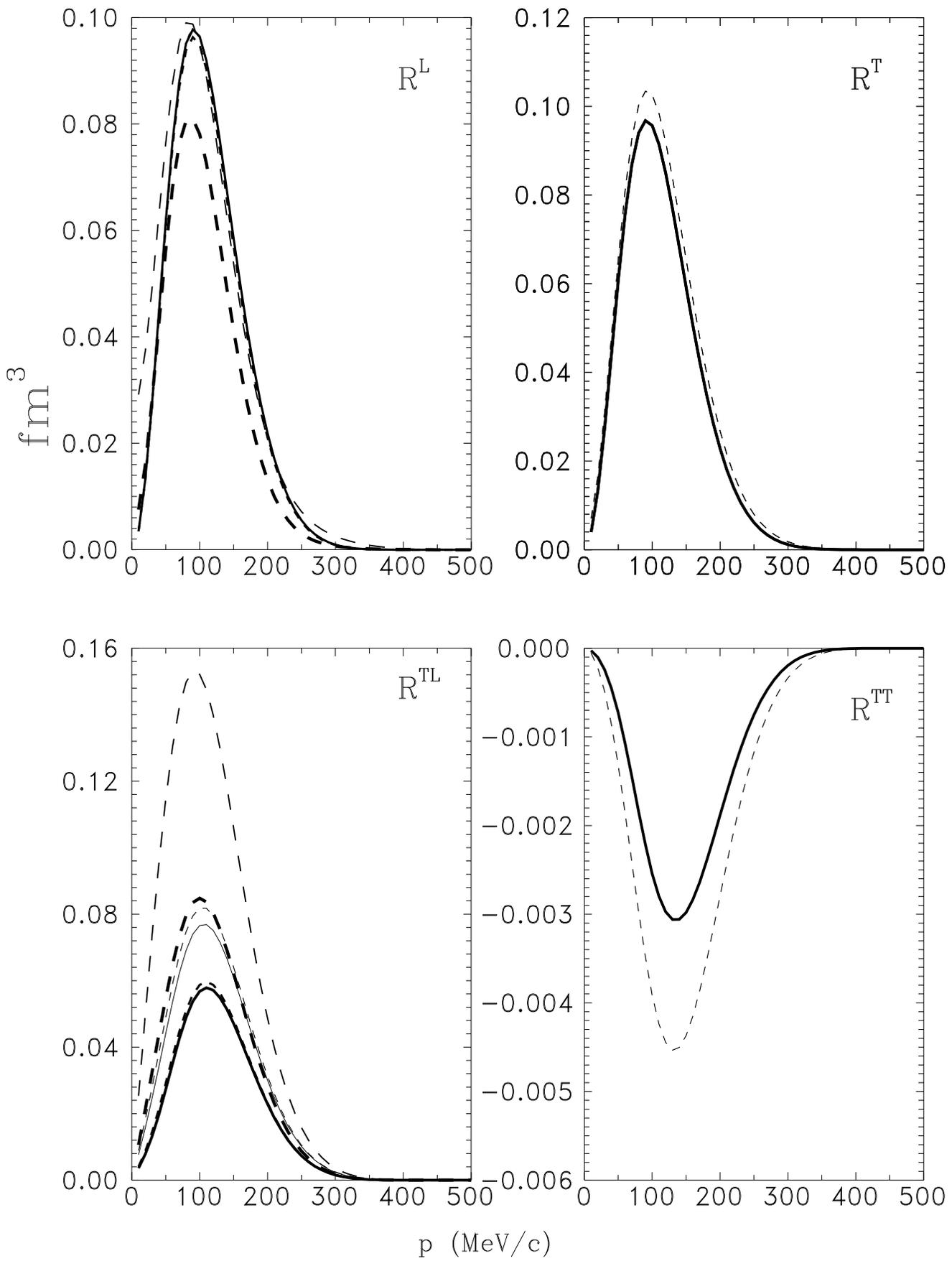
\item[Figure 7:]
Hadronic response functions for the $1p_{1/2}$ shell in $^{16}$O and
kinematics I. The labels of the various curves are as in fig.~2a.
Note that for the two pure transverse responses, 
$R^T$ and $R^{TT}$, all the CC2 (CC1) prescriptions collapse into 
a single solid (dashed) curve.

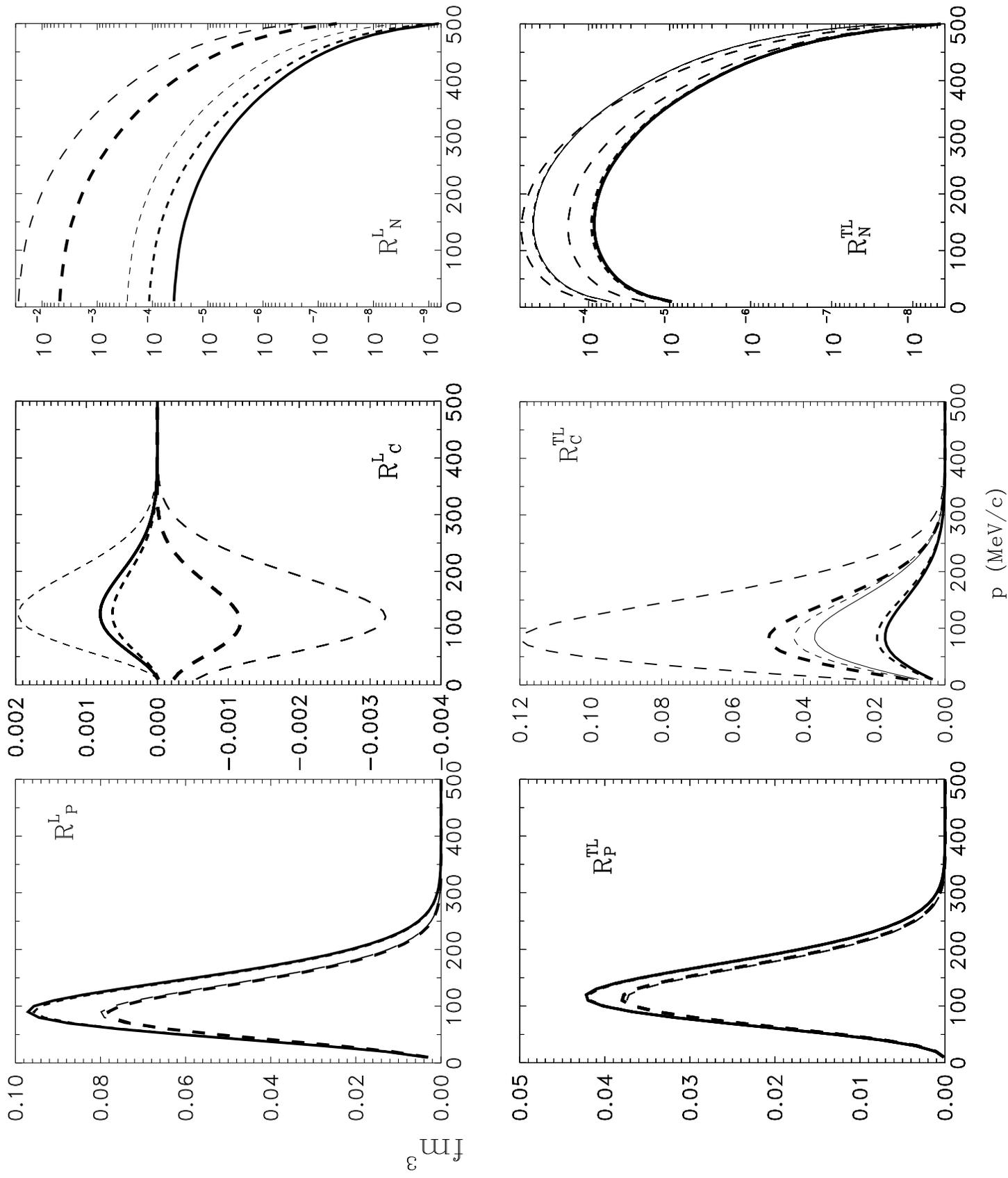
\item[Figure 8:]
Components of the nuclear responses, $R^K_P$, $R^K_C$ and $R^K_N$ 
(see eqs.~(\ref{eqx}--\ref{eq5c})).
Results are shown for longitudinal (top panel) and
tranverse-longitudinal interference (bottom panel) response functions. 
The rest of labeling is as in fig.~2a.

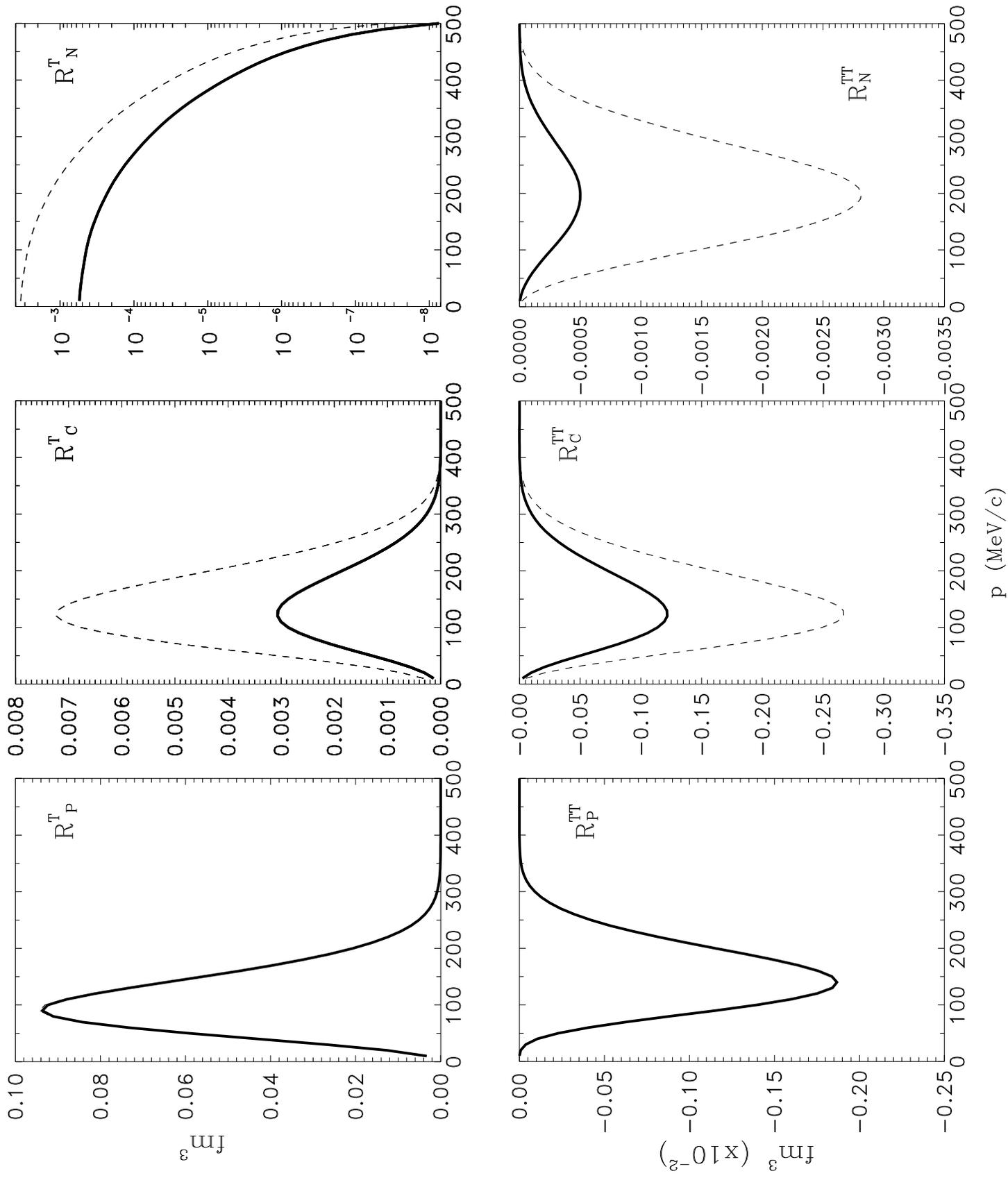
\item[Figure 9:]
Same as fig.~8, except that now for the transverse (top) and
tranverse-transverse interference (bottom) responses.

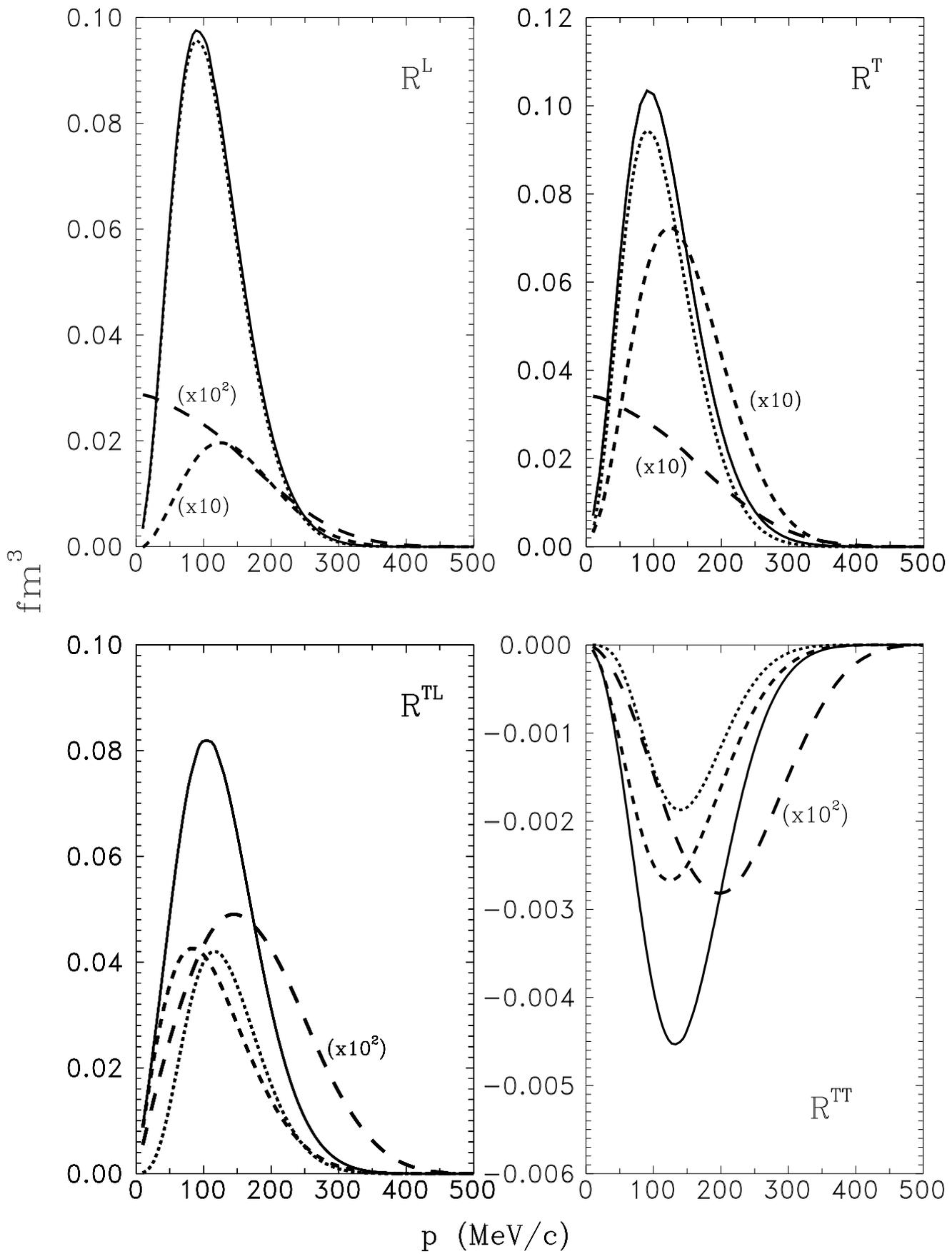
\item[Figure 10:]
Hadronic response functions for the $1p_{1/2}$ shell in $^{16}$O
and kinematics I. All of the results correspond to the
prescription CC1$^{(0)}$. For each response we compare the fully
relativistic result (solid) with the three components as introduced by
eqs.~(\ref{eq5a}--\ref{eq5c}): $R^K_{P}$ (dotted),
$R^K_{C}$ (dashed) and $R^K_{N}$ (long-dash). Some of the
curves are multiplied by different scale factors.

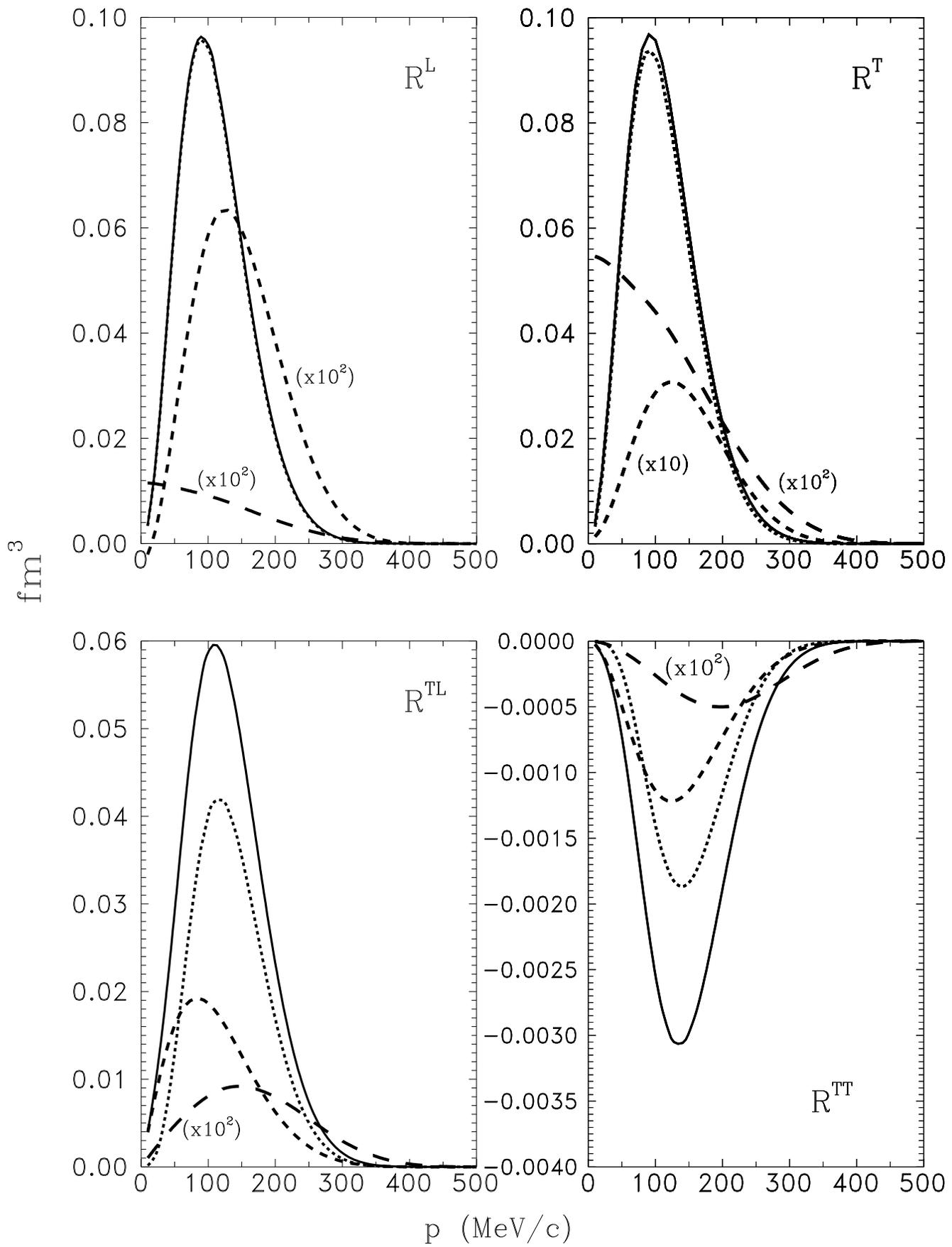
\item[Figure 11:]
Same as fig.~10, except that now the prescription CC2$^{(0)}$ has been
chosen.

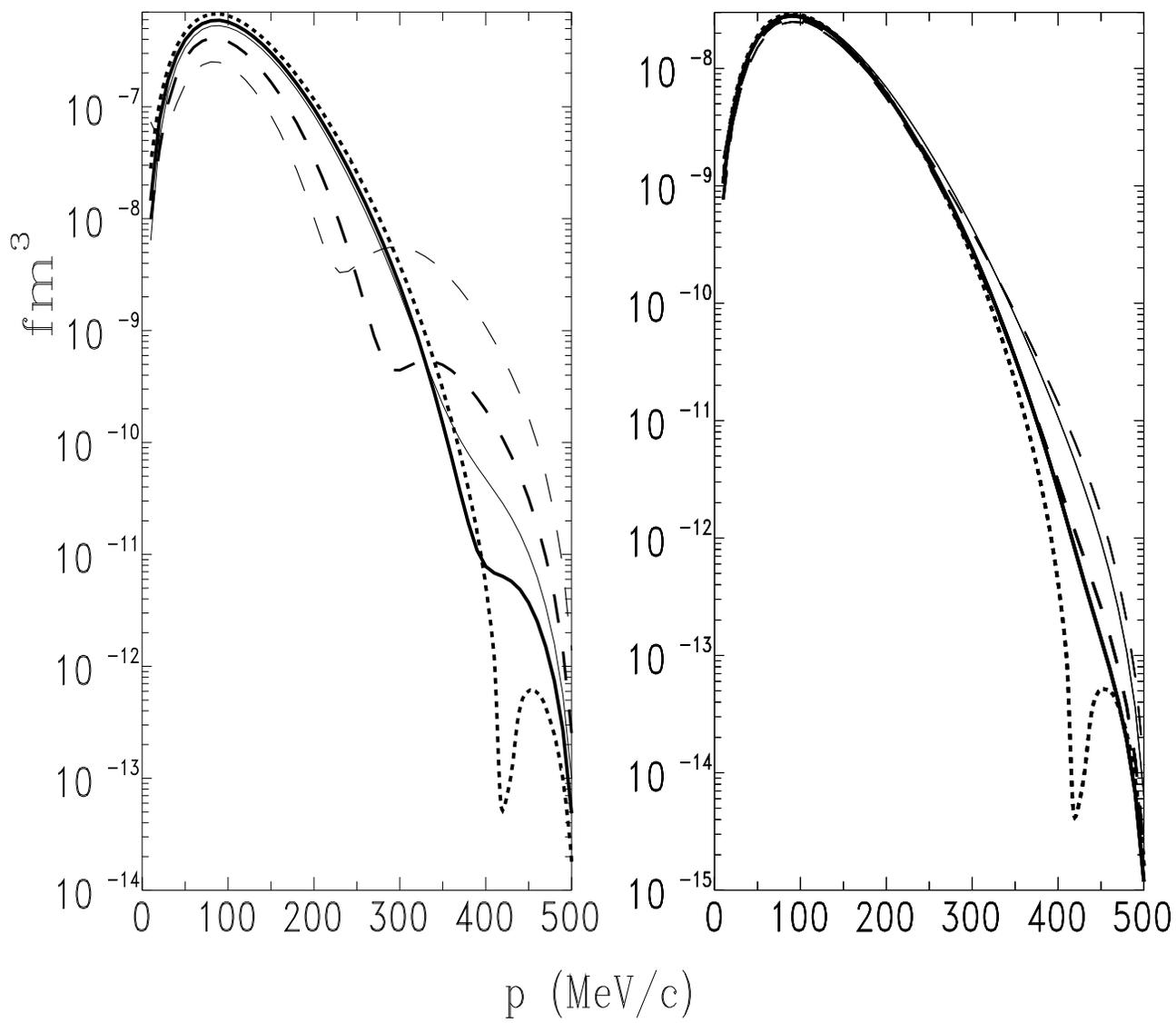
\item[Figure 12:]
Differential cross section for the $1p_{1/2}$ shell in $^{16}$O.
Kinematics I have been chosen and results are presented for forward
and backward electron scattering angles: $\theta_e=30^0$ (left panel) and
$\theta_e=150^0$ (right panel). We show the
fully relativistic results given by the prescriptions (see text):
CC2$^{(0)}$ (thick--solid), CC1$^{(0)}$ (thin--solid), 
CC2$^{(3)}$ (thick--dash) and
CC1$^{(3)}$ (thin--dash). We also show the results corresponding to the
non--relativistic PWIA limit (dotted line) (see text for details).

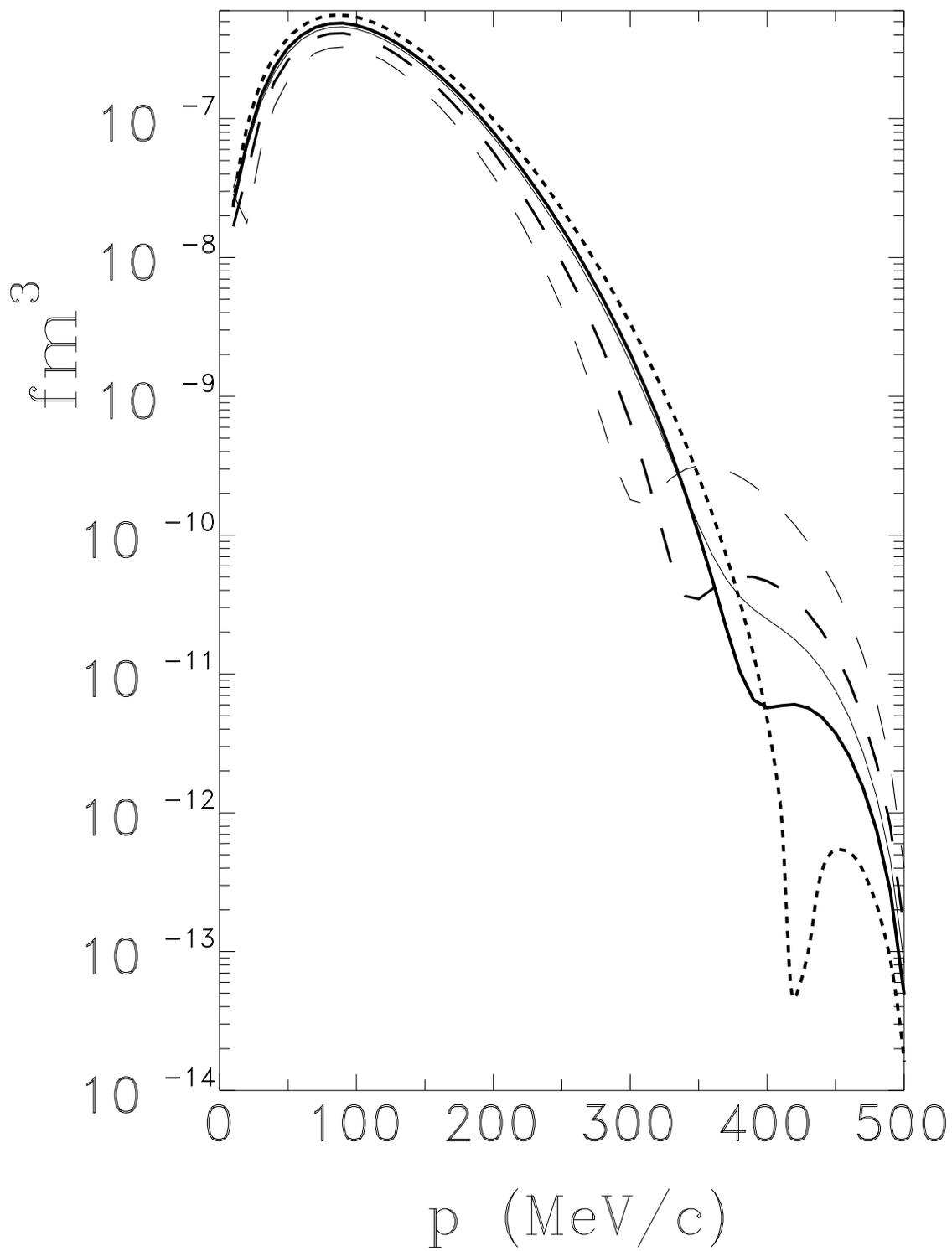
\item[Figure 13:]
Same as fig.~12, except that now for kinematics II and forward
scattering angle ($\theta_e=12.5^0$).

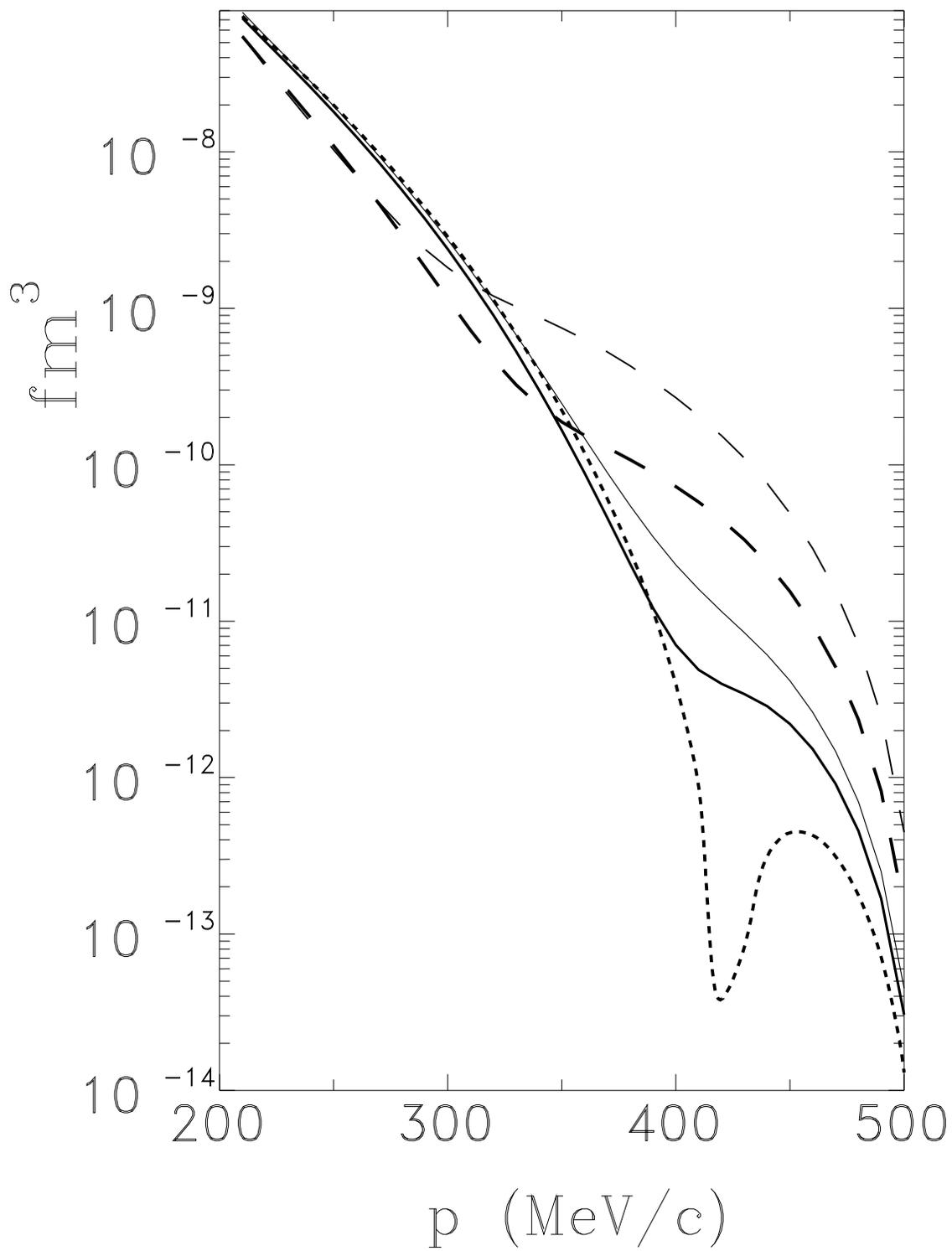
\item[Figure 14:] 
Same as fig.~12, except that now for kinematics III.

\end{enumerate}


\begin{thebibliography}{99}

\bibitem{fm85}  See for instance S. Frullani and J. Mougey, {\bf Adv. Nucl.
Phys.} \underline{14} (1985); T. de Forest, {\bf Nucl. Phys. A }%
\underline{132} (1969) 305; A.E.L. Dieperink and T. de Forest, 
{\bf Ann. Rev. Nucl. Sci.} \underline{25} (1975) 1.

\bibitem{hu90}  P.K.A. de Witt Huberts, {\bf J. Phys. G} \underline{16}
(1990) 507. L. Lapik\'as, {\bf Nucl. Phys. A} \underline{553} (1993) 297C.
J.B. Lanen et al., {\bf Nucl Phys. A} \underline{560} (1993) 811. J.
Wesseling et al., {\bf Phys. Rev. C} \underline{55} (1997) 2773.

\bibitem{qu88}  E.N.M. Quint, Ph. D. Thesis, University of Amsterdam (1988).

\bibitem{bo94}  I. Bobeldijk et al., {\bf Phys. Rev. Lett.} \underline{73}
(1994) 2684.

\bibitem{pa84}  V.R. Pandharipande, C.N. Papanicolas and J. Wambach, {\bf Phys.
Rev. Lett.} \underline{53} (1984) 1133. Z.Y. Ma and J. Wambach, {\bf Phys.
Lett. B} \underline{256} (1991) 1. C. Mahaux and R. Sartor, {\bf Adv. Nucl.
Phys.} \underline{20} (1991) 1.

\bibitem{mu94}  H. M\"uther and W.H. Dickhoff, {\bf Phys. Rev. C} \underline{%
49} (1994) R17.

\bibitem{mc90}  J.P. McDermott, {\bf Phys. Rev. Lett.} \underline{65} (1990)
1991. Y. Jin, D.S. Onley and L.E. Wright, {\bf Phys. Rev. C} \underline{45}
(1992) 1311.

\bibitem{ud93}  J.M. Ud\'\i as, P. Sarriguren, E. Moya de Guerra, E. Garrido
and J.A. Caballero, {\bf Phys. Rev. C} \underline{48} (1993) 2731. J.M.
Ud\'\i as, Ph. D. Thesis, Universidad Aut\'onoma de Madrid (1993).

\bibitem{ud95}  J.M. Ud\'\i as, P. Sarriguren, E. Moya de Guerra, E. Garrido
and J.A. Caballero {\bf Phys. Rev. C} \underline{51}
(1995) 3246.

\bibitem{ud96}  J.M. Ud\'\i as, P. Sarriguren, E. Moya de Guerra, 
and J.A. Caballero {\bf Phys. Rev. C} \underline{53}
(1996) R1488.

\bibitem{ha90}  S. Hama, B.C. Clark, E.D. Cooper, H.S. Sherif and R.L.
Mercer, {\bf Phys. Rev. C} \underline{41} (1990) 2737. E.D. Cooper, S.
Hama, B.C. Clark and R.L. Mercer, {\bf Phys. Rev. C} \underline{47} (1993)
297.

\bibitem{ho91}  C.J. Horowitz, D.P. Murdock and B.D. Serot, in {\bf %
Computational Nuclear Physics} (Eds. K. Langanke, J.A. Maruhn and S.E.
Koonin), Springer-Verlag, Berlin (1991).

\bibitem{se86}  B.D. Serot and J.D. Walecka, {\bf Adv. Nucl. Phys.} 
\underline{16} (1986) 1.

\bibitem{ji94}  Y. Jin and D.S. Onley, {\bf Phys. Rev. C} \underline{45}
(1994) 377. M. Hedayati-Poor, J.I. Johansson and H.S. Sherif, {\bf Phys.
Rev. C} \underline{51} (1995) 2044.

\bibitem{br84}  S.J. Brodsky, {\bf Comm. Nucl. Part. Phys.} \underline{12}
(1984) 213. N. Thies, {\bf Phys. Lett.} \underline{166 B} (1986) 23. E.D.
Cooper and B.K. Jennings, {\bf Nucl. Phys. A} \underline{458} (1986) 717;
G.E. Brown, W. Weise, G. Baym and J. Speth, Comments in Nucl. Part. Phys.
\underline{17} (1987) 39;
S.J. Wallace, F. Gross and J.A. Tjon, {\bf Phys. Rev. Lett.} \underline{74}
(1995) 228.

\bibitem{edu} E. Garrido, J.A. Caballero, E. Moya de Guerra, P. Sarriguren
and J.M. Ud\'{\i}as, Nucl. Phys. {\bf A584} (1995) 256.

\bibitem{bd} J.D. Bjorken and S.D. Drell, {\sc Relativistic Quantum
Mechanics}, McGraw--Hill, N.Y. (1964).

\bibitem{don} A.S. Raskin and T.W. Donnelly, Ann. Phys. {\bf 191}
              (1989) 78.

\bibitem{ann} J.A. Caballero, E. Garrido, E. Moya de Guerra, P. Sarriguren
and J.M. Ud\'{\i}as, Ann. of Phys. {\bf 239} (1995) 351.

\bibitem{dp} J.A. Caballero, T.W. Donnelly, and G. Poulis, 
             Nucl. Phys. {\bf A555} (1993) 709.

\bibitem{fo} T. de Forest, Nucl. Phys. {\bf A392} (1983) 232.

\bibitem{naus} H.W.L. Naus, S.J. Pollock, J.H. Koch and U. Oelfke,
		Nucl. Phys. {\bf A509} (1990) 717;
S. Pollock, H.W.L. Naus and J.H. Koch,
	{Phys. Rev.} {\bf C53} (1996) 2304.

\bibitem{hor} C.J. Horowitz and B.D. Serot, Nucl. Phys. {\bf A368}, (1981)
503; Phys. Lett. B {\bf 86}, (1979) 146.

\bibitem{riska} F. Gross and D.O. Riska, Phys. Rev.
{\bf C36} (1987) 1928.

\bibitem{progress} J.A. Caballero, E. Moya de Guerra, T.W. Donnelly and
J.M. Ud\'{\i}as, {\em work in progress}.

\bibitem{spal93} G.M. Spaltro {\em et al.}, Phys. Rev. C {\bf 48} (1993)
2385. 

\bibitem{bul93} H.J. Bulten, Ph. D. thesis, University of Utrech (1992).
 L. Lapik\'as, Nucl. Phys. A {\bf 553} (1993) 297c.

\bibitem{slu94} V. van der Sluys, J. Ryckebush and M. Waroquier, Phys. Rev. C {\bf 49}
(1994) 2695.

\end{thebibliography}
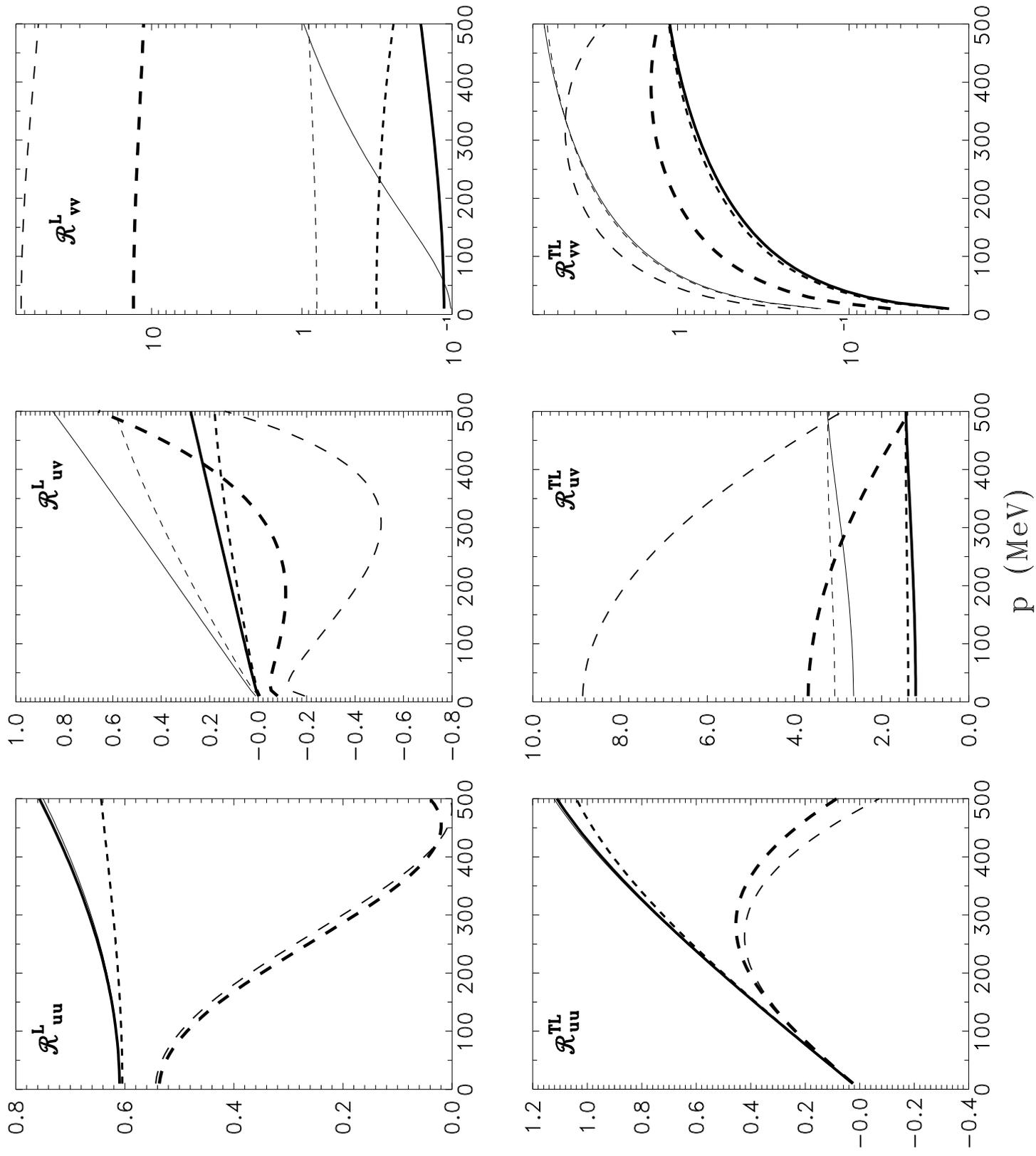
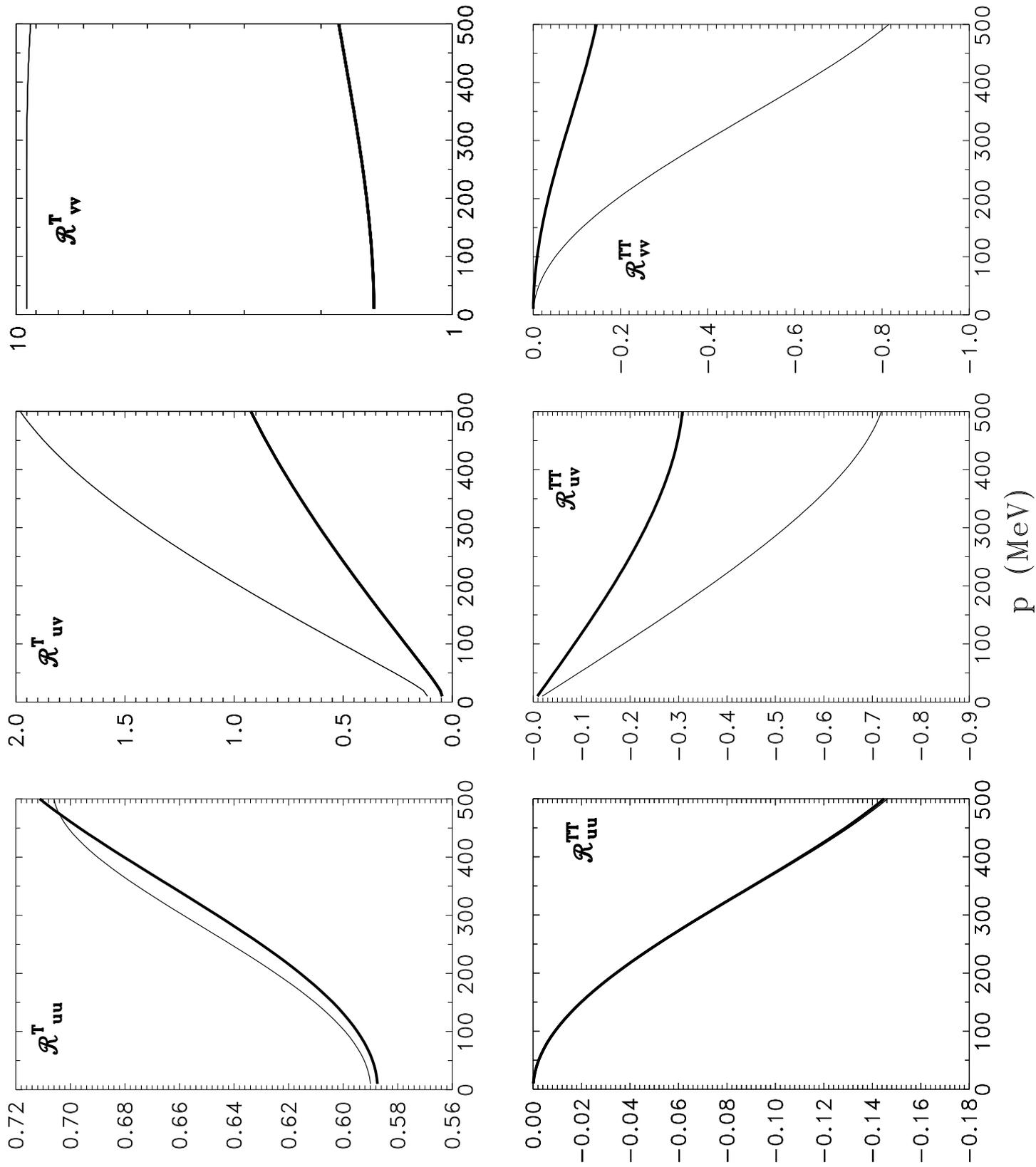
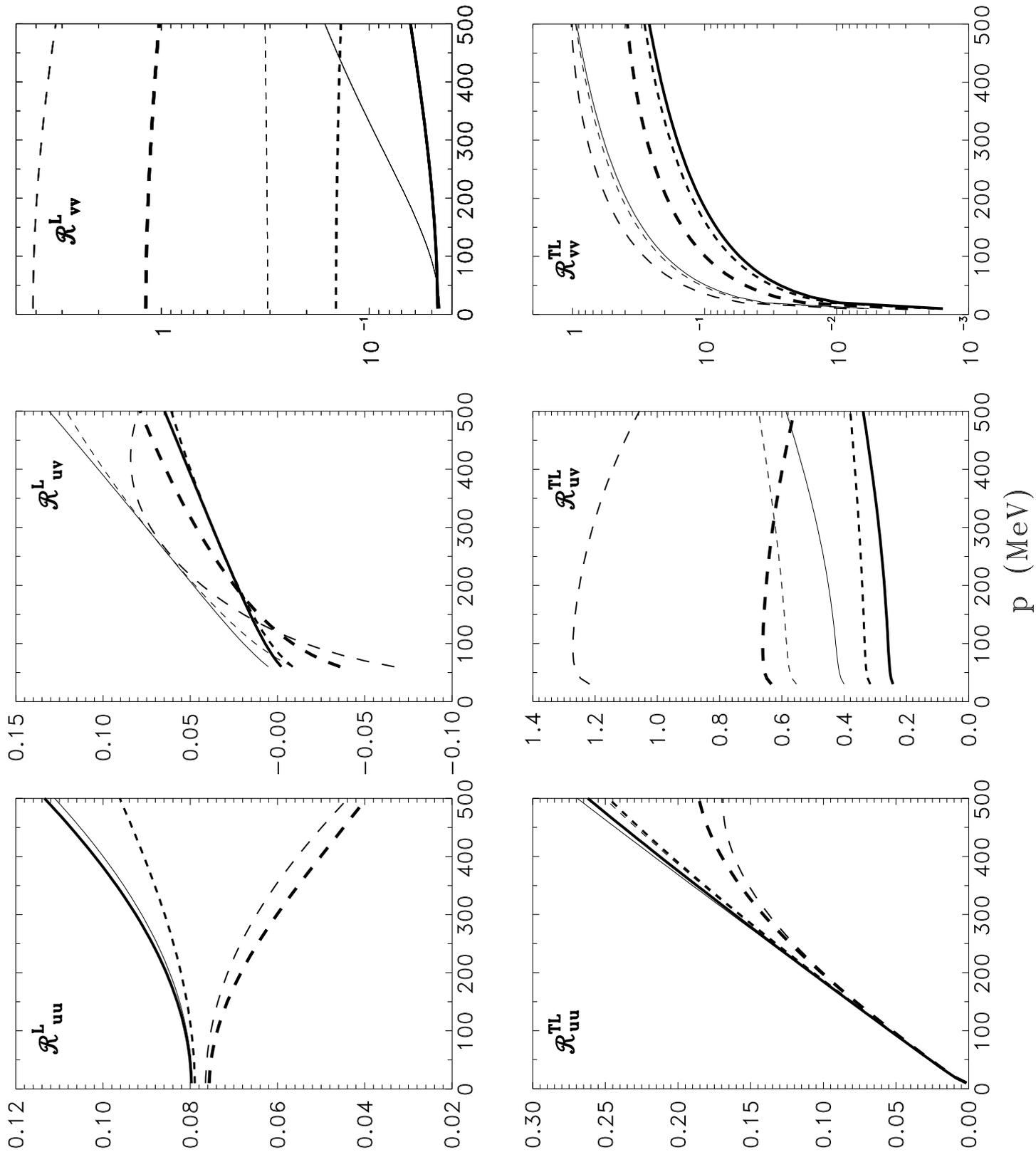
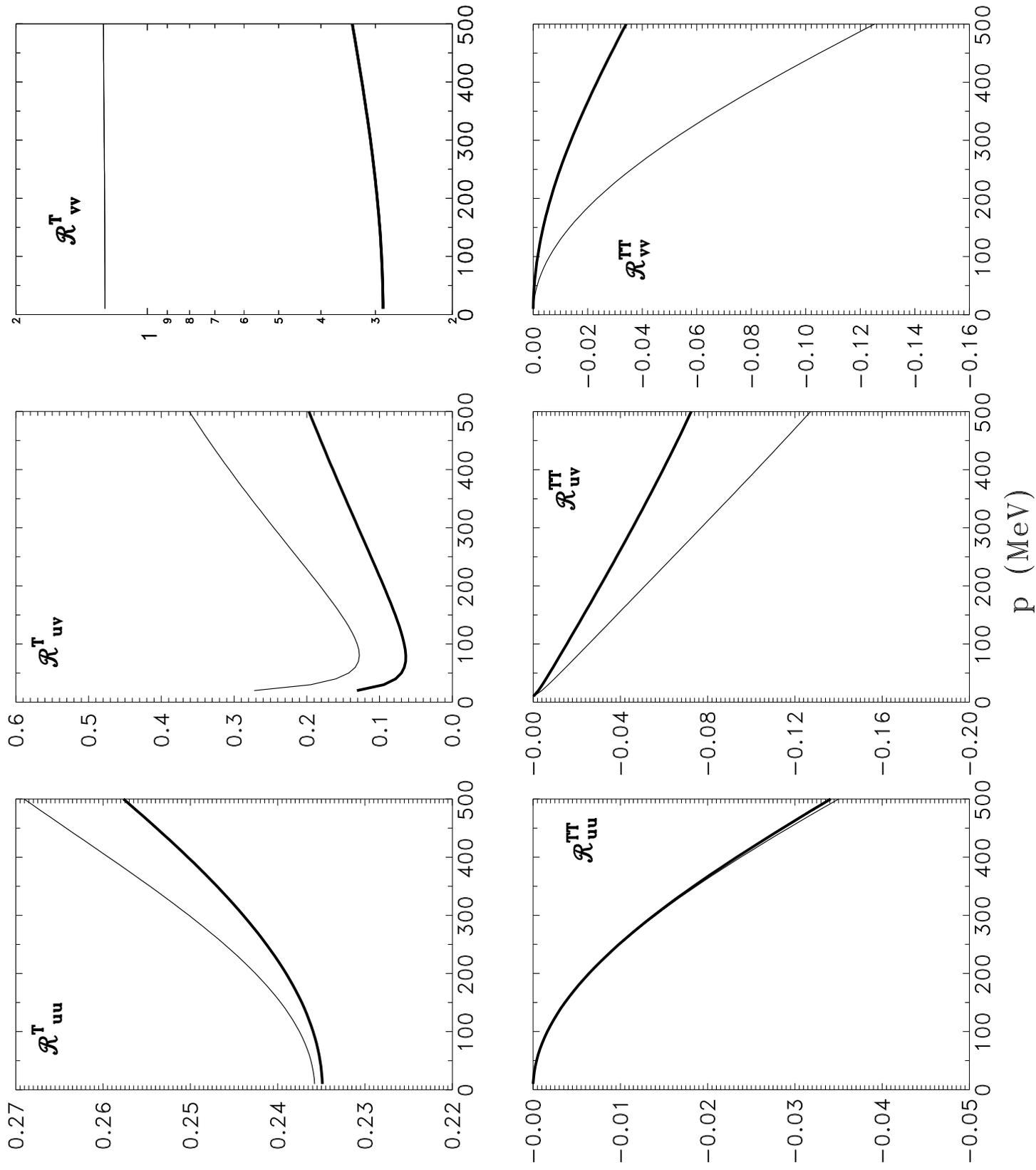
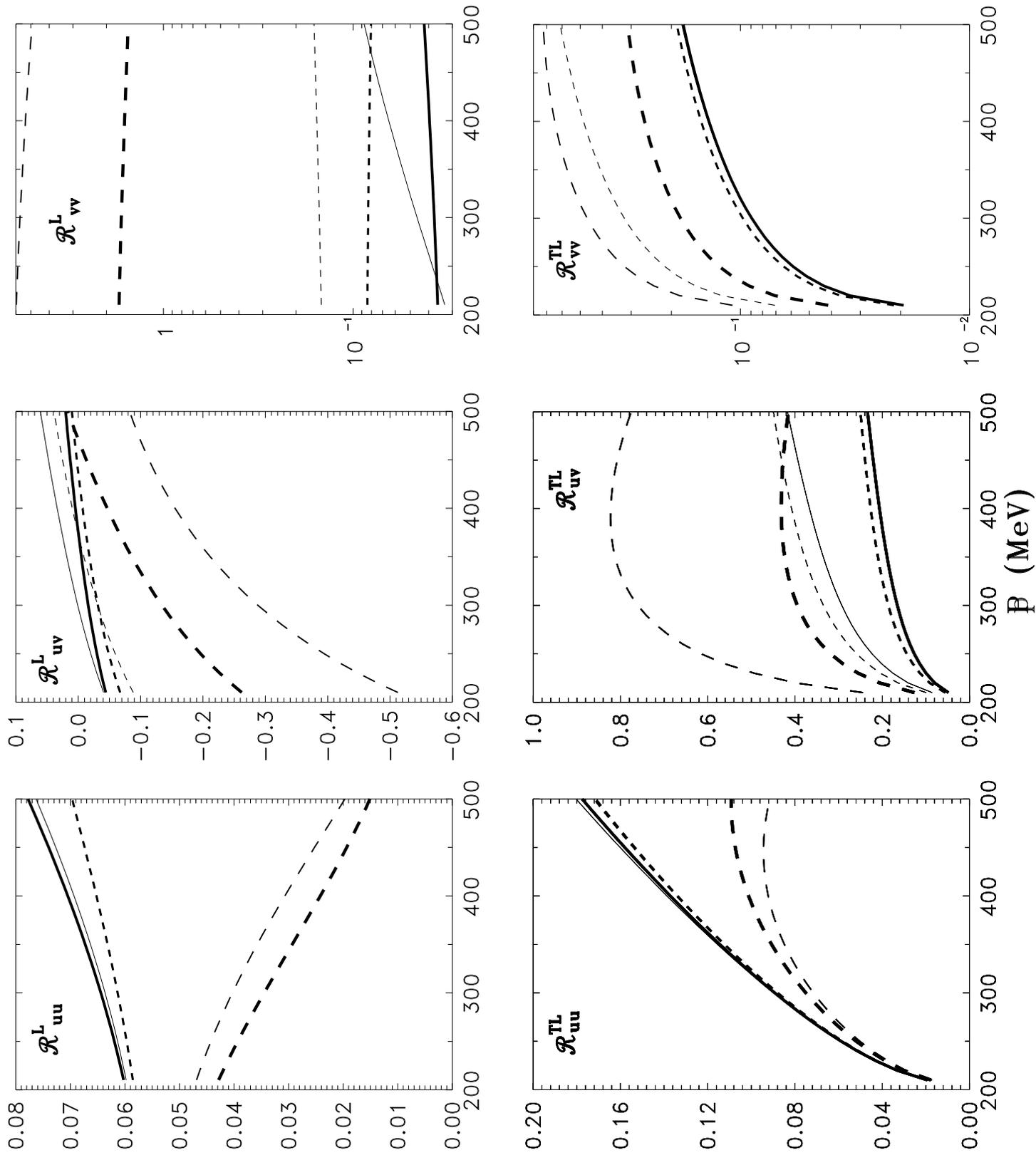
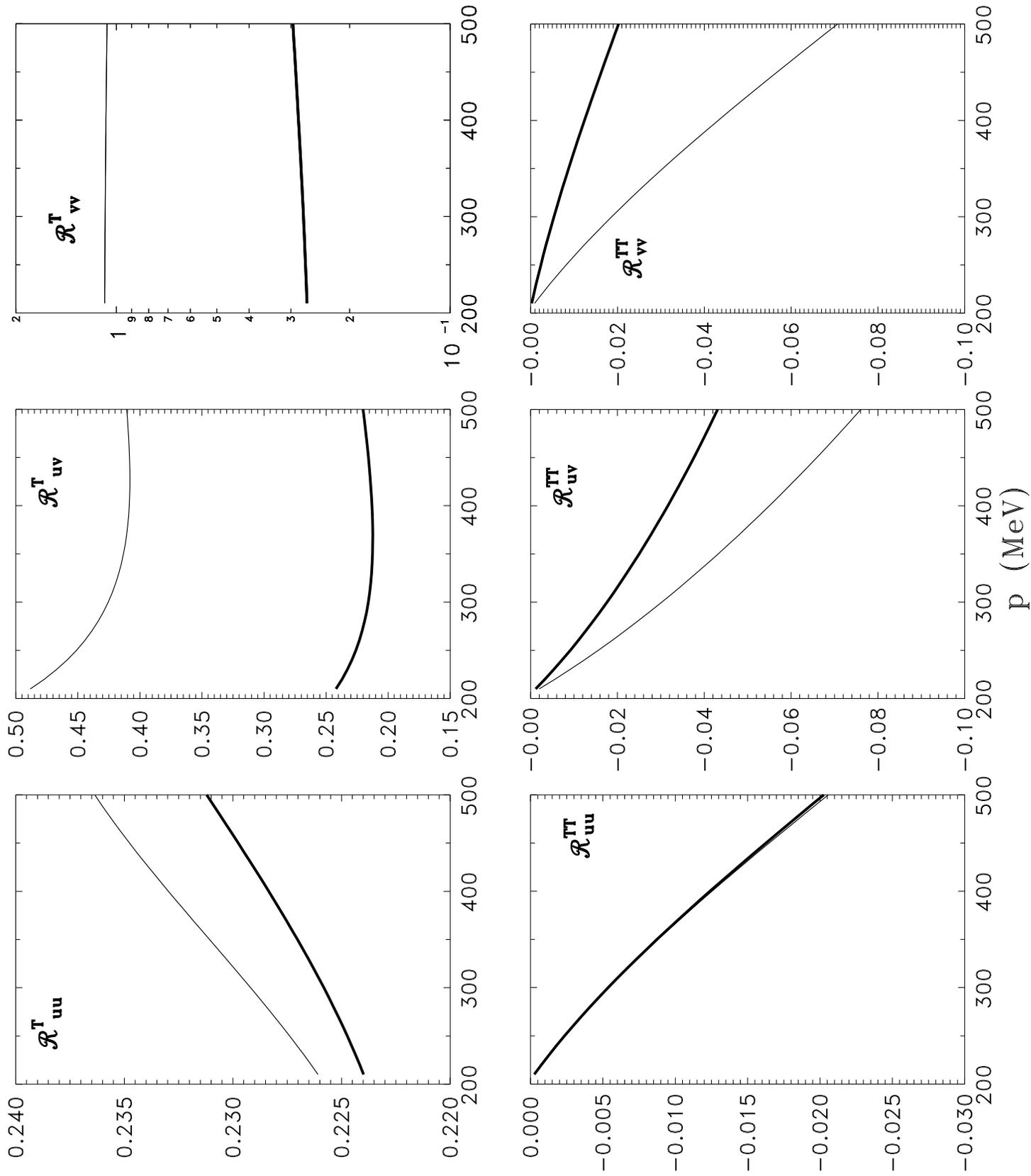
\end{document}